\documentclass[superscriptaddress,aps,prd,singlecolumn,showpacs,nofootinbib,longbibliography, superscriptaddress, showkeys]{revtex4-2}
\DeclareUnicodeCharacter{02BC}{'}
\usepackage[T1]{fontenc}
\usepackage[utf8]{inputenc}  
\DeclareUnicodeCharacter{03B8}{\ensuremath{\theta}}
\DeclareUnicodeCharacter{03B4}{\ensuremath{\delta}}
\usepackage[utf8]{inputenc}
\usepackage{amsmath,amssymb,amsthm}
\usepackage{easybmat}
\usepackage[colorlinks=true,citecolor=blue,urlcolor=blue, linkcolor= magenta]{hyperref}
\usepackage[pdftex]{graphicx}
\usepackage{times, txfonts}
\usepackage{braket}
\usepackage{color}
\usepackage{ragged2e}
\usepackage{booktabs}
\usepackage{epstopdf}
\usepackage{gensymb}
\usepackage{natbib}
\usepackage{subcaption}
\usepackage{float}
\newcommand{\abs}[1]{\left|#1\right|}

\newcommand{\be}{\begin{equation}}
	\newcommand{\ee}{\end{equation}}
\newcommand{\ba}{\begin{eqnarray}}
	\newcommand{\ea}{\end{eqnarray}}

\usepackage{tikz,xcolor,hyperref}

\definecolor{lime}{HTML}{A6CE39}
\DeclareRobustCommand{\orcidicon}{\hspace{-4pt}
	\begin{tikzpicture}
		\draw[lime, fill=lime] (0,0) 
		circle [radius=0.16] 
		node[white] {\hspace{0.1mm}{\fontfamily{qag}\selectfont \tiny ID}};
		\draw[white, fill=white] (-0.07,0.1) 
		circle [radius=0.01];
	\end{tikzpicture}
	\hspace{-3.2mm}
}

\foreach \x in {A, ..., Z}{\expandafter\xdef\csname 
	orcid\x\endcsname{\noexpand\href{https://orcid.org/\csname orcidauthor\x\endcsname}
		{\noexpand\orcidicon}}
}

\begin{document}
 \title{Quantum spread complexity as a probe of NSI, $CP$ Violation, and mass ordering in neutrino oscillations in matter}
	\author{Abhishek Kumar Jha\orcidA{}}\email{kjabhishek@iisc.ac.in}
    \affiliation{Department of Physics, Indian Institute of Science, Bangalore 560012, India}
\author{Govind Krishna G\orcidB{}}\email{m23ph1006@iitj.ac.in }
\author{Anandbhai Pravinbhai Prajapati\orcidC{}}\email{m23ph1002@iitj.ac.in}
\author{Subhashish Banerjee\orcidD{}}\email{subhashish@iitj.ac.in}
\affiliation{Indian Institute of Technology Jodhpur, Jodhpur 342011, India}
	



\begin{abstract}
    {Quantum spread complexity characterizes how a quantum state evolves and becomes distributed over the Hilbert space under unitary dynamics. In this work, we employ a cost function as a quantitative measure of spread complexity. We investigate this cost function within the framework of three-flavor neutrino oscillations in vacuum and matter, incorporating the $CP$-violation phase and Non-Standard Interaction (NSI) effects, under both normal and inverted mass ordering scenarios. The cost function is evaluated for each scenario and analyzed with the corresponding neutrino transition probabilities for both initial muon neutrino and muon antineutrino flavor states. The results are presented using the energy where the first oscillation is maximum and baseline lengths of ongoing long-baseline accelerator neutrino experiments, including T2K and NOvA, as well as upcoming experiments such as DUNE and P2O. Our findings indicate that the difference in the cost function between normal and inverted mass orderings during neutrino propagation in matter is sensitive to these experiments, with the appropriate choice of NSI parameters and the best-fit $CP$-violation phase values.}
\end{abstract}
\maketitle

\section{Introduction}
Neutrino oscillation is a quantum phenomenon in which neutrinos transition between different flavor states ($\nu_e$, $\nu_\mu$, $\nu_\tau$) as they evolve over time \cite{10.1093/acprof:oso/9780198508717.001.0001}. This behavior, observed in various well-established experiments \cite{Super-Kamiokande:1998kpq,kajita2016nobel,mcdonald2016nobel}, provides compelling evidence that neutrinos have non-zero masses and they interact very weakly with matter  \cite{PhysRevD.17.2369,Mikheyev:1985zog}. It also implies that the flavor states are quantum superpositions of non-degenerate mass eigenstates. As these mass eigenstates evolve over
time, the flavor state itself undergoes time evolution and can be expressed as a coherent superposition in the flavor basis. The theoretical foundation of neutrino oscillations was laid by B. Pontecorvo, who first proposed the idea of neutrino flavor conversion analogous to kaon oscillations \cite{Pontecorvo:1957cp,Pontecorvo:1957qd,Pontecorvo:1967fh,BILENKY1978225,Bilenky:2004xm,Bilenky:2016pep}. This idea was later formalized by Maki, Nakagawa, and Sakata, resulting in the now well-known PMNS (Pontecorvo–Maki–Nakagawa–Sakata) matrix \cite{maki1962remarks}. The PMNS matrix for Dirac neutrinos, being unitary in nature, provides the transformation between neutrino flavor and mass eigenstates. It depends on fundamental vacuum parameters such as the mixing angles
($\theta_{12}$, $\theta_{13}$, $\theta_{23}$), mass-squared differences ($\Delta m^2_{21}$, $\Delta m^2_{32}$), and a complex phase $\delta_{\text{CP}}$ linked to charge-conjugation and parity-reversal ($CP$) symmetry violations \cite{Giganti:2017fhf}. In the ultra-relativistic regime, the transition probabilities of initial neutrino flavor states can be related to fundamental vacuum parameters as well as the ratio of length
to energy (L/E), where L is the distance between the neutrino source and detector, and E is the neutrino source energy. Precise measurements of fundamental vacuum parameters \cite{Giganti:2017fhf}, $CP$-violation phases (matter-antimatter asymmetry) \cite{PhysRevD.23.1666,T2K:2019bcf,PhysRevD.57.4403}, effects of Non-Standard Interactions (NSI) \cite{Ohlsson_2013,Chatterjee:2020kkm,Dev_2019}, and unresolved issues such as normal or inverted mass ordering of neutrinos \cite{Kobzarev:1981ra}, and the possible existence of sterile neutrinos \cite{DayaBay:2016lkk,MINOS:2016viw}, are a key area of investigation in current and future neutrino experiments. These include studies using solar \cite{KamLAND:2013rgu}, atmospheric \cite{Denton:2022een}, reactor-based (both short- and long-baseline) \cite{DayaBay:2013yxg,RENO:2018dro,DoubleChooz:2019qbj}, and accelerator-based long-baseline experiments \cite{DiLodovico:2015nfq,Kudryavtsev_2016}.

 The quantum superposition characteristics of neutrino flavor states have been explored by mapping them to mode states, which resemble Bell's like superposition state in the realm of quantum information \cite{Blasone_2009,BLASONE2013320,PhysRevD.77.096002,PhysRevA.77.062304,Banerjee:2015mha,Alok:2014gya,Dixit:2017ron,Dixit:2018kev,Dixit:2019swl,Dixit:2020ize,KumarJha:2020pke,Li:2022mus,wang2023monogamy,Bittencourt:2023asd,Blasone:2023qqf,Quinta:2022sgq,Jha:2021itm,Jha:2020qea,Jha:2022yik,Konwar:2024nrd,Alok:2024xeg,Alok:2025qqr,Dixit:2019lsg,Jha:2024gdq,Jha:2024asl,Jha:2025ekn}. Owing to their weak interaction with matter, neutrino beams can preserve quantum coherence over macroscopic distances, a property that holds potential relevance for developments in quantum information. Along these lines, using the plane wave approach, various quantum correlations have been investigated by quantifying them in terms of neutrino transition probabilities \cite{Blasone_2009,BLASONE2013320,PhysRevD.77.096002,PhysRevA.77.062304,Banerjee:2015mha,Alok:2014gya,Dixit:2017ron,Dixit:2018kev,Dixit:2019swl,Dixit:2020ize,KumarJha:2020pke,Li:2022mus,wang2023monogamy,Bittencourt:2023asd,Blasone:2023qqf,Quinta:2022sgq,Jha:2021itm,Jha:2020qea,Jha:2022yik,Konwar:2024nrd,Alok:2024xeg,Alok:2025qqr,Dixit:2019lsg,Jha:2024gdq,Jha:2024asl,Jha:2025ekn}. These include concurrence, linear entropy, teleportation fidelity, geometric discord, von Neumann entropy, tangle, negativity, three-tangle, three-$\pi$, as well as other nonclassical features such as violations of the Bell and Bell-CHSH (Clauser-Horne-Shimony-Holt) inequalities, among others. Efforts have been made to analyze some of these quantum correlations within the framework of experimentally observed neutrino oscillations \cite{PhysRevA.98.050302,Dixit:2018gjc,Li:2022zic,Yadav:2022grk,Konwar:2024pkh,Bouri:2024kcl}. The study has also been extended using the wave-packet approach \cite{Blasone_2015,Blasone:2021cau,Blasone:2022ete,Ravari:2022yfd,Ettefaghi:2023zsh}. These studies have provided clues that multimode entanglement persists during neutrino oscillations. Furthermore, the non-classical nature of neutrino oscillations has been explored through temporal analogues of Bell inequalities, particularly Leggett-Garg inequalities \cite{PhysRevLett.117.050402,PhysRevD.99.095001,NAIKOO2020114872,Blasone:2022iwf,Chattopadhyay:2023xwr,Soni:2023njf,Shafaq:2020sqo,Sarkar:2020vob,Konwar:2024nwc}. In the present work, we explore an alternative measure of neutrino transition probabilities, namely the quantum spread complexity in three-flavor neutrino oscillations.

Quantum spread complexity is a concept originating from quantum information theory and quantum dynamics, reflecting how a quantum state evolves and spreads
over a Krylov basis as the system evolves \cite{PhysRevD.106.046007,Haque:2022ncl}. It quantifies how ‘complex’ the state becomes in terms of its distribution across the eigenstates of a driving Hamiltonian. In more concrete terms, if a
quantum state initially localized in one basis state spreads out over many basis states under time evolution, its spread complexity increases. This growth of complexity can be quantitatively measured by the cost function \cite{PhysRevD.106.046007}.  A cost function quantifies the ‘cost’ or ‘distance’ associated with quantum operations, which define the geometry of the space of quantum states or unitary transformations. Minimizing such cost functions corresponds to finding
optimal paths in the Hilbert space that transform one quantum state
into another with minimal ‘effort’ or complexity. In recent years, the concept of quantum spread complexity has been investigated in various areas of physics, including black holes \cite{Susskind:2014moa,Susskind:2014rva,Stanford:2014jda,Brown:2015bva,Brown:2015lvg}, quantum chaos \cite{Ali:2019zcj,Bhattacharyya:2019txx,Dymarsky:2019elm,Bhattacharyya:2020iic,Balasubramanian:2019wgd,Bhattacharyya:2020art,Balasubramanian:2021mxo}, quantum phase transitions \cite{Ali:2018aon,PhysRevB.106.195125,Caputa_2023}, decoherence in open systems \cite{Bhattacharyya:2022rhm,Bhattacharyya:2021fii}, PT-symmetric transitions \cite{Beetar:2023mfn,Bhattacharya:2024hto}, integrability breaking transitions \cite{Balasubramanian:2024ghv,Camargo:2024deu}, cosmology \cite{Bhattacharyya:2020rpy,Bhattacharyya:2020kgu,Haque:2021hyw}, quantum computation \cite{Nielsen:2006cea,Nielsen:2005mkt,Dowling:2006tnk,Jefferson:2017sdb,Ali:2018fcz}, high-energy physics \cite{Parker:2018yvk,Pedraza:2022dqi,Bhattacharjee:2022vlt}, and other dynamical phenomena such as neutrino oscillation \cite{Dixit:2023fke}.

From a quantum dynamical perspective, neutrino oscillation is effectively described by a time-dependent quantum state evolving in a finite-dimensional Hilbert space spanned by flavor or mass eigenstates. The state of a neutrino initially localized in one flavor spreads over the different flavor basis as it propagates, naturally
leading to the idea of spread complexity. The spread complexity can quantify how the neutrino state evolves from an initial flavor into a
superposition of multiple flavor states. By defining suitable cost functions associated with the unitary transformations driving neutrino flavor evolution, typically governed by the PMNS mixing matrix and using the corresponding driving Hamiltonian, one can characterize the cost function in experimentally observed neutrino oscillations. This approach offers a novel viewpoint: neutrino oscillation phenomena, beyond their standard probabilistic interpretation, can be understood in terms of complexity growth in the neutrino flavor quantum state space. This interplay has implications in understanding fundamental questions such
as how quantum complexity develops in naturally occurring quantum systems such as neutrinos, and how the structure of neutrino mixing relates to optimal paths or minimal cost transformations in flavor space. Recent studies have begun exploring these
connections \cite{Dixit:2023fke}, suggesting that complexity measures could potentially serve as tools for analyzing neutrino oscillations and their sensitivity to $CP$-violation phases or matter interactions. In the present work, we investigate quantum spread complexity in three-flavor neutrino oscillations in vacuum and matter, incorporating NSI and $CP$-violation phase under both normal and inverted mass ordering scenarios. We characterize our results for both the initial muon neutrino and muon antineutrino flavor states using the fundamental vacuum parameters of the NuFIT data \cite{Esteban:2024eli,NuFIT}, along with the baseline lengths and energies of ongoing long-baseline accelerator neutrino experiments such as T2K (Tokai to Kamioka, Japan) and NOvA (NuMI Off-axis $\nu_e$ Appearance, USA), as well as the upcoming DUNE (Deep Underground Neutrino Experiment, USA) and P2O (Protvino to ORCA, Russia) experiments.

The organization of the paper is as follows: In Sec.\,\ref{Sec2}, we provide a brief introduction to quantum spread complexity and the associated cost function. In Sec.\,\ref{Sec3}, we discuss the formalism of three-flavor neutrino oscillations in vacuum, a constant matter potential, and NSI, incorporating $CP$-violation phases. In Sec.\,\ref{Sec4}, quantum spread complexity and the cost function are computed in the three-flavor framework, with a particular focus on the time evolution of the initial muon neutrino and muon
antineutrino flavor states propagating through vacuum, a constant matter potential, and in the presence of NSI, under both normal and inverted mass ordering scenario, with their respective $CP$-violation phase. Finally, Sec.\ref{Sec5} provides a discussion of the results and conclusions.

\section{Quantum spread complexity and cost function}
\label{Sec2}
For a general quantum state $|\psi(t)\rangle$, the time evolution can be written using Schrodinger equation as
\begin{equation}\label{1}
    \iota \frac{\partial}{\partial t}|\psi(t)\rangle=H|\psi(t)\rangle.
\end{equation}
The solution gives the time evolution, which is
\begin{equation}\label{2}
    |\psi(t\rangle)=e^{-\iota Ht}|\psi(0)\rangle.
\end{equation}
The quantum spread complexity can be defined as the spread of the state $|\psi (t)\rangle$ in the Hilbert space relative to the initial state $|\psi(0)\rangle$, where $|\psi (t)\rangle$ is the target state and $|\psi(0)\rangle$ is the reference state related to each other by unitary transformations~\cite{PhysRevD.106.046007, PhysRevB.106.195125}.
 Eq.\,(\ref{2}) can be written in a series form given by \cite{PhysRevD.106.046007}
\begin{equation}\label{3}
    |\psi(t)\rangle = \sum_{n=0}^{\infty}\frac{(-\iota t)^{n}}{n!}H^{n}|\psi(0)\rangle=\sum_{n=0}^{\infty}\frac{(-\iota t)^{n}}{n!}|\psi_{n}\rangle,
\end{equation}
where 
\begin{equation}\label{3a}
    |\psi_{n}\rangle=H^{n}|\psi(0)\rangle.
\end{equation}
The evolved state $|\psi(t)\rangle$ is written as a superposition of infinite $|\psi_{n}\rangle$ states. The set of states: $\{|\psi_{0}\rangle,|\psi_{1}\rangle,|\psi_{2}\rangle,...\}$ are not necessarily orthonormal. However, to construct the Krylov basis we require orthonormal set of states. Therefore, we use the Gram-Schmidt procedure to make orthonormal set of states and convert them into Krylov basis. Here
\begin{align}\label{4a}
    &|K_{0}'\rangle=|\psi(0)\rangle,&\nonumber\\
   &|K_{1}'\rangle=|\psi(1)\rangle-\frac{\langle K_{0}'|\psi_{1}\rangle}{\langle K_{0}'|K_{0}'\rangle}|K_{0}'\rangle,&\nonumber\\
 &|K_{2}'\rangle=|\psi(2)\rangle-\frac{\langle K_{0}'|\psi_{2}\rangle}{\langle K_{0}'|K_{0}'\rangle}|K_{0}'\rangle-\frac{\langle K_{1}'|\psi_{2}\rangle}{\langle K_{1}'|K_{1}'\rangle}|K_{1}'\rangle,
\end{align}
and so on. These states are made orthonormal by the following procedure
\begin{align}\label{4}
    |K_{0}\rangle=\frac{|K_{0}'\rangle}{\sqrt{\langle K_{0}'|K_{0}'\rangle}},\nonumber\\
    |K_{1}\rangle=\frac{|K_{1}'\rangle}{\sqrt{\langle K_{1}'|K_{1}'\rangle}},\nonumber\\
    |K_{2}\rangle=\frac{|K_{2}'\rangle}{\sqrt{\langle K_{2}'|K_{2}'\rangle}}.
\end{align}
The set of orthonormal vectors $\{|K_{0}\rangle,|K_{1}\rangle,|K_{2}\rangle,...\}$ forms the Krylov basis, $\{|K_{n}\rangle\}$~\cite{PhysRevD.106.046007}. Depending on the choice of initial state and the dynamics involved, the elements of the Krylov basis may have fewer elements than the dimension of Hilbert space. As time evolution becomes more and more complex, the spread of the evolved state in the given Hilbert space also increases. The cost function is defined as a measure of this spread complexity from a minimum of all possible basis choices~\cite{PhysRevD.106.046007}. As the Krylov basis holds all the relevant information about the system with the least number of basis elements, it can be used to define the cost function. The cost function can be expressed as this~\cite{PhysRevD.106.046007, Caputa_2023} 
\begin{equation}\label{5}
    \chi=\sum C_{n}|\langle K_{n}|\psi(t)\rangle|^{2}=\sum C_{n}P_{K_{n}} ,
\end{equation}
where $C_{n}$ is a real increasing number such as $C_{n}=n=0,1,2,...$ and $P_{K_{n}}$ is the probability of $\psi(t)$ being in one of the Krylov basis.

\section{Three Flavor Neutrino Oscillation in Presence of NSI}
\label{Sec3}
In the three-flavor neutrino oscillation, at time $t=0$, the mixing between the flavor eigenstates $|\nu_{\alpha}\rangle$ ($\alpha=e,\mu,\tau$) and mass eigenstates $|\nu_{i}\rangle$ ($i=1,2,3$) is given by \cite{10.1093/acprof:oso/9780198508717.001.0001}
\begin{equation}\label{6}
    |\nu_{\alpha}\rangle=U^{\star}|\nu_{i}\rangle,
\end{equation}
where $U$ is the Pontecorvo-Maki-Nakagawa-Sakata (PMNS) mixing matrix ~\cite{maki1962remarks}, and the asterisk ($*$) denotes the complex conjugate of $U$. Here, $U$\footnote{If neutrinos are Majorana particles, the PMNS matrix contains two additional complex phases; however, these phases do not impact neutrino flavor transition probabilities \cite{Giganti:2017fhf}.} is defined as \cite{maki1962remarks,Giganti:2017fhf}
\begin{equation}\label{7}
    U=\begin{bmatrix} 
U_{e1} & U_{e2} & U_{e3} \\  U_{\mu1} & U_{\mu2} & U_{\mu3} \\  U_{\tau1} & U_{\tau2} & U_{\tau3}
\end{bmatrix}=
\begin{bmatrix}
    C_{12}C_{13} & S_{12}C_{13} & 
    S_{13}e^{-\iota \delta_{\text{CP}}}\\
    -S_{12}C_{23}-C_{12}S_{23}S_{13}e^{\iota \delta_{\text{CP}}} & C_{12}C_{23}-S_{12}S_{23}S_{13}e^{\iota \delta_{\text{CP}}} & S_{23}C_{13}
    \\ S_{12}S_{23}-C_{12}C_{23}S_{13}e^{\iota \delta_{\text{CP}}} & -C_{12}S_{23}-S_{12}C_{23}S_{13}e^{\iota \delta_{CP}} & C_{23}C_{13}
\end{bmatrix},
\end{equation}
where $C_{ij}=\cos{\theta_{ij}}$, $S_{ij}=\sin{\theta_{ij}}$ with $\theta_{ij}$ as mixing angles and $\delta_{\text{CP}}$ is the $CP$-violation phase. Here, we discuss the time evolution of three-flavor neutrino oscillations in vacuum, a constant matter potential, and in the presence of NSI parameters, with $\delta_{\text{CP}}$ in both normal and inverted mass ordering scenarios. We follow the description provided in Ref.\,\cite{Dixit:2023fke}

The Schrodinger equation represents the evolution of the flavor eigenstates as\footnote{Throughout this work, we adopt natural units by setting $\hbar = c = 1$.}
\begin{equation}\label{8a}
    i\frac{\partial}{\partial t}\begin{bmatrix}
      |\nu_{e}(t)\rangle\\
      |\nu_{\mu}(t)\rangle\\
      |\nu_{\tau}(t)\rangle
    \end{bmatrix}=\mathcal{H}_{\text{total}}\begin{bmatrix}
      |\nu_{e}(t)\rangle\\
      |\nu_{\mu}(t)\rangle\\
      |\nu_{\tau}(t)\rangle
    \end{bmatrix},
\end{equation}
where $\mathcal{H}_{\text{total}}$ is the effective Hamiltonian in flavor basis $\ket{\nu_\alpha}$, defined as \cite{10.1093/acprof:oso/9780198508717.001.0001,Chatterjee:2020kkm,Konwar:2024nrd}
\begin{equation}\label{11}
    \mathcal{H}_{\text{total}}=U{\mathcal{H}}_{\text{vac}}U^\dagger + \mathcal{\mathcal{H}}_{\text{mat}}+ \mathcal{H}_{\text{NSI}}.
\end{equation}
Here, $\mathcal{H}_{\text{vac}}$ is the Hamiltonian in the mass basis $\ket{\nu_i}$, represented as \cite{10.1093/acprof:oso/9780198508717.001.0001}
\begin{equation}\label{12}
  \mathcal{H}_{\text{vac}} = \frac{1}{2E}\begin{bmatrix}
        0 & 0 & 0\\
        0 &  \Delta m^{2}_{21} & 0 \\
        0 & 0 & \Delta m^{2}_{31} 
    \end{bmatrix},
\end{equation}
where $\Delta m_{21}^2=m^2_{2}-m^2_{1} $ and $ \Delta m_{31}^2=m^2_3-m^2_1$  are the neutrino mass-squared differences, and \( E \) is the neutrino energy which is different for different neutrino experiments. $\mathcal{H}_{\text{mat}}$ and $\mathcal{H}_{\text{NSI}}$ are the Hamiltonians in flavor basis when neutrinos oscillate in matter and the presence of non-standard interactions (NSI), respectively. The plane wave solution of Eq.\,(\ref{8a}) can be interpreted as the evolution of the flavor eigenstate in the superposition of the flavor basis, expressed as
 \begin{align}\label{9}
    \begin{bmatrix}
      |\nu_{e}(t)\rangle\\
      |\nu_{\mu}(t)\rangle\\
      |\nu_{\tau}(t)\rangle
    \end{bmatrix}&=e^{-\iota \mathcal{H}_{\text{total}}t}\begin{bmatrix}
      |\nu_{e}\rangle\\
      |\nu_{\mu}\rangle\\
      |\nu_{\tau}\rangle
    \end{bmatrix}&\nonumber\\
   & =\begin{bmatrix}
        A_{ee}(t) & A_{e\mu}(t) & A_{e\tau}(t) \\
        A_{\mu e}(t) & A_{\mu\mu}(t) & A_{\mu\tau}(t) \\
        A_{\tau e}(t) & A_{\tau \mu}(t) & A_{\tau\tau}(t)
    \end{bmatrix}\begin{bmatrix}
      |\nu_{e}\rangle\\
      |\nu_{\mu}\rangle\\
      |\nu_{\tau}\rangle
    \end{bmatrix}.
\end{align}
In the Standard Model, neutrinos interact only via weak interaction, which can be mediated through either charged currents (CC) or neutral currents (NC)\footnote{The neutral current (NC) term is omitted from $\mathcal{H}_{\text{mat}}$ in Eq.\,(\ref{13}) since it introduces a common phase to all neutrino flavors, which can be factored out through a phase shift and does not impact the neutrino flavor transition probabilities \cite{10.1093/acprof:oso/9780198508717.001.0001}.}.
When neutrinos travel through the Earth-matter background they interact with electrons via charge current potential ($V_{CC}$), which stems from coherent forward scattering of neutrinos with electrons in matter. The effective Hamiltonian in the presence of a matter potential is defined as ~\cite{PhysRevD.17.2369,Mikheyev:1985zog,Ohlsson_2000}
\begin{equation}\label{13}
   \mathcal{H}_{\text{mat}} = 
    \begin{pmatrix}
    V_{CC} & 0 & 0 \\
    0 & 0 & 0 \\
    0 & 0 & 0
    \end{pmatrix},
    \quad
    V_{CC} = \pm \sqrt{2}G_{F}N_{e}
\end{equation}
where $G_F$ is the Fermi constant and $N_e$ is the electron number density.
 The sign of matter potential depends on whether it is a neutrino or an antineutrino. For neutrino, the matter potential is positive ($+V_{CC}$), and for antineutrino, the matter potential is negative ($-V_{CC}$). Corresponding to a matter density of 2.8 gm/cc, the matter potential is approximated to be $V_{CC}\approx1.01\times 10^{-13}\,\text{eV}$ \cite{Bouri:2024kcl}.

Moreover, NSI are hypothetical interactions beyond the Standard Model that can affect neutrino oscillations. These interactions can be manifested through four-fermion dimension-6 operators, affecting charged-current (CC) and neutral-current (NC) interactions~\cite{Davidson_2003, Antusch_2009, Ohlsson_2013, farzan2018neutrinooscillationsnonstandardinteractions, Dev_2019}. As NC-NSI has relatively milder constraints than highly constrained CC-NSI~\cite{Konwar:2024nrd}, we can exclude CC-NSI effects in our calculations. The NC Lagrangian can be given by \cite{Konwar:2024nrd}
\begin{equation}\label{14}
    \mathcal{L}_{\text{NSI}}^{\text{NC}}=2\sqrt{2}G_{F}\sum_{\alpha,\beta,P}\epsilon_{\alpha \beta}^{f,P}(\bar{\nu}_{\alpha}\gamma^{\mu}P\nu_{\beta})(\bar{f}\gamma_{\mu}Pf),
\end{equation}
where $\alpha$ and $\beta$ correspond to different neutrino flavors and $P\in\{P_{L}, P_{R}\}$,$P_{L, R}=(1\mp \gamma^{5})/2$ are the left handed and right handed chirality operators, respectively. $\epsilon_{\alpha\beta}^{f, P}$ is a dimensionless coefficient, which measures the strength of NSI compared to the weak interaction coupling constant $G_{F}$. $f\in\{e,u,d\}$ denotes the matter fields. The Hamiltonian in the presence of NSI can be written as \cite{Chatterjee:2020kkm,Konwar:2024nrd}
\begin{equation}\label{15}
    \mathcal{H}_{\text{NSI}}= V_{CC}\begin{bmatrix}
        \epsilon_{ee}(x) & \epsilon_{e\mu}(x) & \epsilon_{e\tau}(x)\\
        \epsilon_{\mu e}(x) & \epsilon_{\mu\mu}(x) & \epsilon_{\mu\tau}(x)\\
        \epsilon_{\tau e}(x) & \epsilon_{\tau \mu}(x) & \epsilon_{\tau\tau}(x)
    \end{bmatrix}.
\end{equation}
\begin{table}[H]
\centering
\begin{tabular}{|c|c|c|}
\hline
NSI parameters & Range (1$\sigma$) & Range (2$\sigma$) \\
\hline
$\epsilon^{\oplus}_{ee}$ & $[-0.30, 0.20] \oplus [0.95, 1.3]$ & $[-1.00, 1.4]$ \\
\hline
$\epsilon^{\oplus}_{e\mu}$ & $[-0.12, 0.011]$ & $[-0.20, 0.09]$ \\
\hline
$\epsilon^{\oplus}_{e\tau}$ & $[-0.16, 0.083]$ & $[-0.24, 0.30]$ \\
\hline
$\epsilon^{\oplus}_{\mu\mu}$ & $[-0.43, 0.14] \oplus [0.91, 1.3]$ & $[-0.80, 1.4]$ \\
\hline
$\epsilon^{\oplus}_{\mu\tau}$ & $[-0.047, 0.012]$ & $[-0.021, 0.021]$ \\
\hline
$\epsilon^{\oplus}_{\tau\tau}$ & $[-0.43, 0.21] \oplus [0.83, 1.3]$ & $[-0.85, 1.4]$ \\
\hline
\end{tabular}
\caption{The real values of the NSI parameters are taken from Refs.\,\cite{coloma2023globalconstraintsnonstandardneutrino,Konwar:2024nrd}, and their upper limits, along with the corresponding $2\sigma$ errors, are considered in this work.}
\label{tab:1}
\end{table}

\begin{table}[H]
\centering
\begin{tabular}{|c|c|c|}
\hline
Parameters & Best fit $\pm 1\sigma$ (NO) & Best fit $\pm 1\sigma$ (IO) \\
\hline
$\theta_{12}^\circ$ & $33.68^{+0.73}_{-0.70}$ & $33.68^{+0.73}_{-0.70}$ \\
\hline
$\theta_{13}^\circ$ & $8.52^{+0.11}_{-0.11}$ & $8.58^{+0.11}_{-0.12}$ \\
\hline
$\theta_{23}^\circ$ & $48.5^{+0.7}_{-0.9}$ & $48.6^{+0.9}_{-1.0}$ \\
\hline
$\Delta m^2_{21} \times 10^{-5}~\mathrm{(eV^2)}$ & $7.49^{+0.19}_{-0.19}$ & $7.49^{+0.19}_{-0.19}$ \\
\hline
$\Delta m^2_{3l} \times 10^{-3}~\mathrm{(eV^2)}$ & $+2.534^{+0.025}_{-0.023}$ & $-2.510^{+0.024}_{-0.025}$ \\
\hline
$\delta_{CP}^\circ$ & $177^{+19}_{-20}$ & $285^{+25}_{-28}$ \\
\hline
\end{tabular}
\caption{The fundamental vacuum parameters for NO and IO, considered in this work, are taken from Refs.\,\cite{Esteban:2024eli,NuFIT} along with their corresponding $1\sigma$ errors ($90\%$CL).}
\label{tab:2}
\end{table}

The NSI parameters $\epsilon_{\alpha\beta}(x)$, where $\alpha,\beta=e,\mu,\tau$, can be expressed as
\begin{equation}\label{16}
    \epsilon_{\alpha\beta}(x)=\sum\limits_{f=e,u,d}\frac{N_{f}(x)}{N_{e}(x)}\epsilon^{f}_{\alpha \beta}~.
\end{equation}
Here, $x$ denotes the position or location of the medium through which neutrinos are passing, $f \in e,u,d$ are fermions present in matter, and $N_{f}(x)$ is the density of fermions in matter. From the electric charge neutrality condition $(N_{p}=N_e)$ and the quark structure of proton $(N_{p}=2N_u+N_d)$ and neutron $(N_n=N_u+2N_d)$. Substituting these conditions into Eq.\,\eqref{16}, $\epsilon_{\alpha \beta}(x)$ becomes
\begin{equation}\label{17}
    \epsilon_{\alpha \beta}(x)=\epsilon^e_{\alpha\beta}+(2+Y_n(x))\epsilon_{\alpha \beta}^u+(1+2Y_n(x))\epsilon_{\alpha\beta}^d
\end{equation}
where $Y_n(x)=N_n(x)/N_e(x)$. NSI parameters can be real or complex. For complex off-diagonal parameters, from the Hermitian property of the Hamiltonian we can write $\epsilon_{\alpha\beta }=\epsilon_{\beta\alpha}^*$. If the parameters are real, then $\epsilon_{\alpha\beta }=\epsilon_{\beta\alpha}$ and the Hamiltonian is symmetric in nature. NSI can exhibit both vector(V) and axial-vector (A) types, with $\epsilon_{\alpha\beta}^f=\epsilon_{\alpha\beta}^{f, L}\pm \epsilon_{\alpha\beta}^{f, R}$, where the sign plus ($+$) corresponds to the vector component and the sign minus ($-$) corresponds to the axial-vector component. The vector and axial-vector types of NSI contribute to neutrino oscillations in different ways. The allowed values and constraints on the NSI parameters are derived from analyses of data from a wide range of experiments ~\cite{Esteban_2018, Esteban_2019, Coloma_2020, coloma2023globalconstraintsnonstandardneutrino}. The real values of the NSI parameters are summarized in Table \ref{tab:1}.

For the long-baseline accelerator neutrino experiment, we choose our initial state to be muon flavor state $|\nu_{\mu}\rangle$. Using Eq.\,(\ref{9}), the time evolution of such a state can be written in a linear superposition of flavor basis as
\begin{equation}\label{17a}
    \ket{\nu_\mu(t)}=A_{\mu e}(t) \ket{\nu_e}+ A_{\mu\mu}(t)\ket{\nu_\mu} + A_{\mu\tau}(t)\ket{\nu_\tau},
\end{equation}
where $|A_{\mu e}|^2 +|A_{\mu \mu}|^2+ |A_{\mu \tau}|^2=1$. In the ultra-relativistic regime ($t\approx L$), where $L$ is the distance traveled by the neutrino from source to detector, using Eq.\,(\ref{9}), the three different transition probabilities\footnote{The various transition probability expressions given in Eq.\,(\ref{18}) contain the oscillatory term $\sin^2(\frac{\Delta m^2 L}{4E})$. In the analysis of neutrino oscillation experimental data, this term is commonly expressed in a more convenient form by restoring physical constants: $\sin^2(\frac{\Delta m^2 L}{4E})\equiv\sin^2(\frac{\Delta m^2 Lc^3}{4\hbar E}) $ $\rightarrow \sin^2(1.27\Delta m^2 (\text{eV}^2)\frac{L(\text{km})}{E(\text{GeV})})$ \cite{10.1093/acprof:oso/9780198508717.001.0001}.} of the initial state $\ket{\nu_\mu}$ can be obtained as a function of the ratio $L/E$ as
\begin{equation}{}\label{18}
    P_{\mu \rightarrow e}(L/E)=|A_{\mu e }(L/E)|^2,\hspace{0.5cm} 
     P_{\mu \rightarrow \mu}(L/E)=|A_{\mu\mu}(L/E)|^2, \hspace{0.5cm}
    P_{\mu \rightarrow \tau}(L/E)=|A_{\mu\tau}(L/E)|^2. 
\end{equation}
\begin{table}[H]
\centering
\begin{tabular}{|c|c|c|}
\hline
Experiments & Baseline length ($L$) & Range of Energy ($E$) \\
\hline
T2K & 295\,\text{km} & 0.2  to 1.2
$\text{GeV}$ \cite{NA61_SHINE_T2K_yields_2016}\\
\hline
NOvA & 810\,\text{km} & 1  to 5 $\text{GeV}$ \cite{NOvA:2021nfi} \\
\hline
DUNE & 1300\,\text{km} & 1  to 10 $\text{GeV}$  \cite{DUNE:2015lol,Bouri:2024kcl}\\ 
\hline
P2O & 2595\,\text{km} & 2  to 10 $\text{GeV}$ \cite{Akindinov_2019_LetterOfInterest_P2O}\\
\hline
\end{tabular}
\caption{The expected baseline lengths and energy ranges of ongoing long-baseline accelerator neutrino experiments such as T2K and NOvA, as well as upcoming experiments like DUNE and P2O, considered in this work, are shown in the Table.}
\label{tab:3}
\end{table}
\begin{table}[ht]
\centering
\begin{tabular}{|l|c|c|c|c|}
\hline
Experiments & Vacuum ($\delta_{cp}=0^\circ$) & Matter ($\delta_{cp}=0^\circ$) & Matter ($\delta_{cp} = 177^\circ$) & NSI ($\delta_{cp} = 177^\circ$) \\
\hline
&$E(\text{GeV})$&$E(\text{GeV})$&$E(\text{GeV})$&$E(\text{GeV})$\\
\hline
T2K   & 0.672 & 0.655 & 0.549 & 0.545 \\
NOvA  & 1.80  & 1.70  & 1.44  & 1.47  \\
DUNE  & 2.91  & 2.60  & 2.23  & 2.34  \\
P2O  & 5.82   & 4.65  & 4.05   & 5.10  \\
\hline
\end{tabular}
\caption{\justifying{In the NO scenario, the energy $E\,(\text{GeV})$ corresponding to the first oscillation maximum of the initial muon flavor neutrino state $\ket{\nu_\mu}$ in vacuum, matter, and NSI, with two distinct $\delta_{\text{CP}}$ values. These results are obtained from the transition probability $P_{\mu\rightarrow e}$ across various neutrino experiments such as T2K, NOvA, DUNE, and P2O, as shown in Fig.\,\ref{fig:prob-vs-energy-1}.}}
\label{tab:4}
\end{table}
\begin{table}[ht]
\centering
\begin{tabular}{|l|c|c|c|c|}
\hline
Experiments & Vacuum ($\delta_{cp}=0^\circ$)& Matter ($\delta_{cp}=0^\circ$) & Matter ($\delta_{cp} = 285^\circ$) & NSI ($\delta_{cp} = 285^\circ$) \\
\hline
&$E(\text{GeV})$&$E(\text{GeV})$&$E(\text{GeV})$&$E(\text{GeV})$\\
\hline
T2K   & 0.55  & 0.56 & 0.65 & 0.65 \\
NOvA  & 1.49  & 1.63  & 1.89  & 1.84 \\
DUNE  & 2.41   & 2.75   & 3.21  & 3.07  \\
P2O  & 4.65   & 6.14  & 7.38  & 6.18 \\
\hline
\end{tabular}
\caption{\justifying{In the IO scenario, the energy $E(\text{eV})$ corresponding to the first oscillation maximum of the initial muon flavor neutrino state $\ket{\nu_\mu}$ in vacuum, matter, and NSI, with two distinct $\delta_{\text{CP}}$ values. These results are obtained from the transition probability $P_{\mu\rightarrow e}$ across various neutrino experiments such as T2K, NOvA, DUNE, and P2O, as shown in Fig.\,\ref{fig:prob-vs-energy-2}.}}
\label{tab:5}
\end{table}

Furthermore, determining the ordering of the neutrino mass remains a fundamental challenge, strongly constrained by current data on neutrino transition probabilities, which admit two distinct mass-ordering scenarios \cite{Giganti:2017fhf}. In the normal ordering (NO) framework, $m_3 > m_2 > m_1$. In contrast, the inverted ordering (IO) scenario is $m_3 < m_1 < m_2$. The transition probabilities, as given in Eq.\,(\ref{18}), depend on the fundamental vacuum parameters: $\Delta m^2_{21}$, $\Delta m^2_{31}$ (where $\Delta m_{31}^2=\Delta m_{32}^{2}+\Delta m_{21}^{2}$), $\theta_{12}$, $\theta_{13}$, $\theta_{23}$, and $\delta_{\rm CP}$, along with their best fit $1\sigma$ values for both NO and IO orderings, are summarized in Table\,\ref{tab:2}. It should be noted that the mass-squared difference is defined as $\Delta m^2_{3l} = \Delta m^2_{31} > 0$ for NO, and $\Delta m^2_{3l} = \Delta m^2_{32} < 0$ for IO \cite{Esteban:2024eli,NuFIT}.

In addition to the fundamental vacuum parameters, a constant matter potential, and the NSI parameters, the transition probabilities of the initial state $\ket{\nu_{\mu}}$, as given in Eq.\,(\ref{18}), also depend on the ratio $L/E\,\text{(km/GeV)}$. This ratio can be precisely controlled in various ongoing long-baseline accelerator neutrino experiments such as T2K and NOvA, as well as in upcoming experiments like DUNE and P2O, thereby offering a practical and controlled framework for studying neutrino oscillations. The expected baseline lengths and energy ranges for these experiments are provided in Table\,\ref{tab:3}.

Moreover, in our analysis, we consider different cases under both the NO and IO scenarios. In the NO scenario, the cases include neutrino oscillation in vacuum with $\delta_{\text{CP}}=0$, in a constant matter potential with $\delta_{\text{CP}}=0$ and the best fit $\delta_{\text{CP}}=177^\circ$, and in the presence of NSI with best fit $\delta_{\text{CP}}=177^\circ$. Similarly, in the IO scenario, we consider oscillation in vacuum with $\delta_{\text{CP}}=0$,  in a constant matter potential with $\delta_{\text{CP}}=0$ and best fit $\delta_{\text{CP}}=285^\circ$, and in the presence of NSI with best fit $\delta_{\text{CP}}=285^\circ$.

\begin{figure}[H]
    \centering
    \begin{subfigure}[b]{0.33\textwidth}
        \centering
        \includegraphics[width=\textwidth]{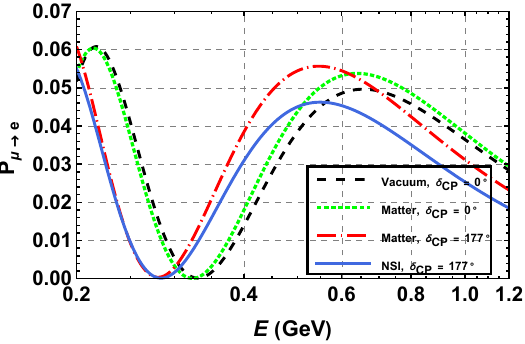}
        \caption{T2K}
        \label{fig:pe1a}
    \end{subfigure}
    \hfill
    \begin{subfigure}[b]{0.33\textwidth}
        \centering
        \includegraphics[width=\textwidth]{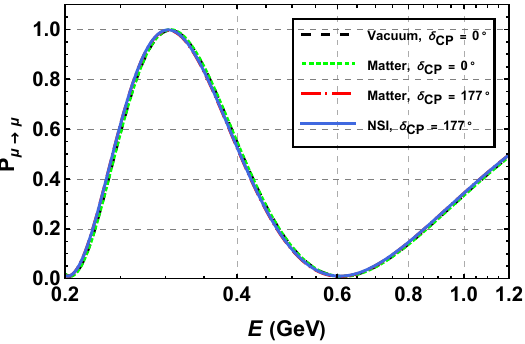}
        \caption{T2K}
        \label{fig:pe1b}
    \end{subfigure}
    \hfill
    \begin{subfigure}[b]{0.33\textwidth}
        \centering
        \includegraphics[width=\textwidth]{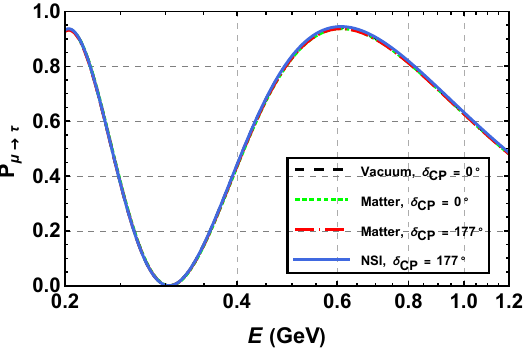}
        \caption{T2K}
        \label{fig:pe1c}
    \end{subfigure}
    \hfill
    \begin{subfigure}[b]{0.33\textwidth}
        \centering
        \includegraphics[width=\textwidth]{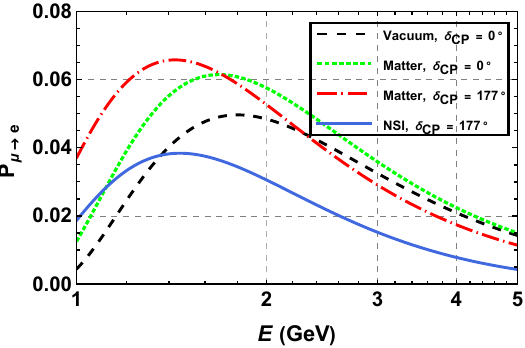}
        \caption{NOvA}
        \label{fig:pe1d}
    \end{subfigure}
    \hfill
     \begin{subfigure}[b]{0.33\textwidth}
        \centering
        \includegraphics[width=\textwidth]{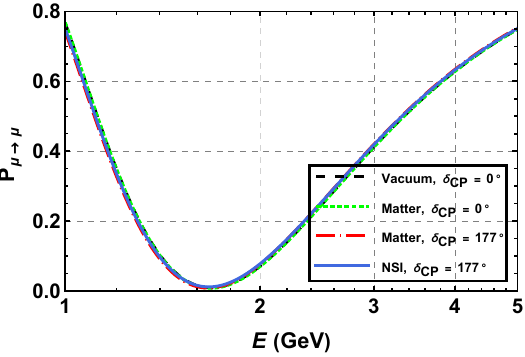}
        \caption{NOvA}
        \label{fig:pe1e}
    \end{subfigure}
    \hfill
    \begin{subfigure}[b]{0.33\textwidth}
        \centering
        \includegraphics[width=\textwidth]{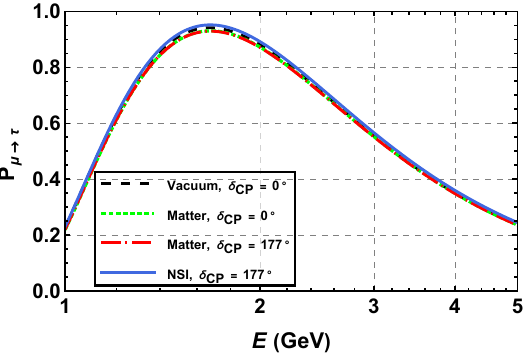}
        \caption{NOvA}
        \label{fig:pe1f}
        \end{subfigure}
         \begin{subfigure}[b]{0.33\textwidth}
        \centering
        \includegraphics[width=\textwidth]{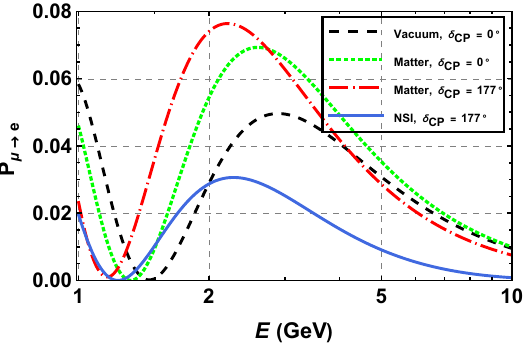}
        \caption{DUNE}
        \label{fig:pe1g}
    \end{subfigure}
    \hfill
    \begin{subfigure}[b]{0.33\textwidth}
        \centering
        \includegraphics[width=\textwidth]{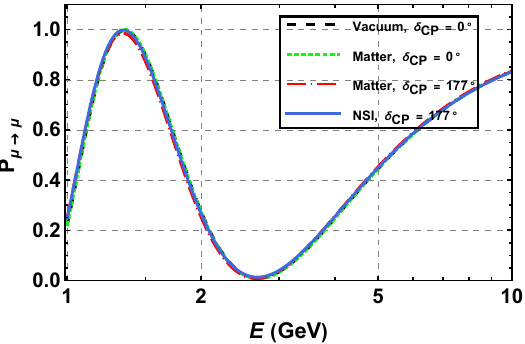}
        \caption{DUNE}
        \label{fig:pe1h}
    \end{subfigure}
    \hfill
    \begin{subfigure}[b]{0.33\textwidth}
        \centering
        \includegraphics[width=\textwidth]{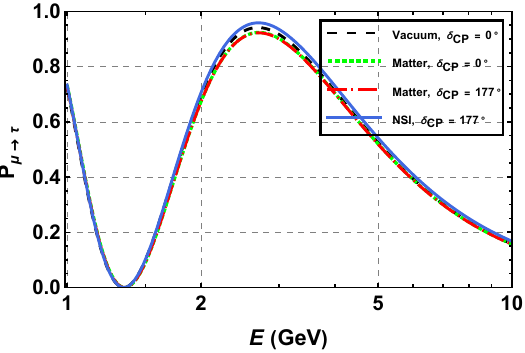}
        \caption{DUNE}
        \label{fig:pe1i}
    \end{subfigure}
    \begin{subfigure}[b]{0.33\textwidth}
        \centering
        \includegraphics[width=\textwidth]{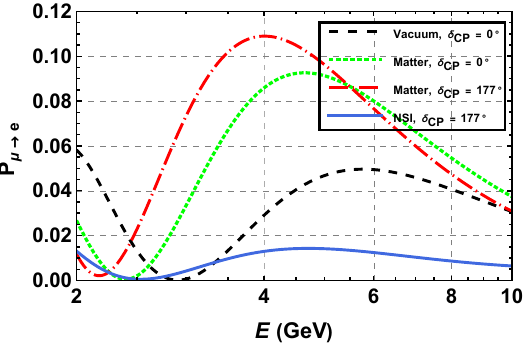}
        \caption{P2O}
        \label{fig:pe1j}
    \end{subfigure}
    \hfill
    \begin{subfigure}[b]{0.33\textwidth}
        \centering
        \includegraphics[width=\textwidth]{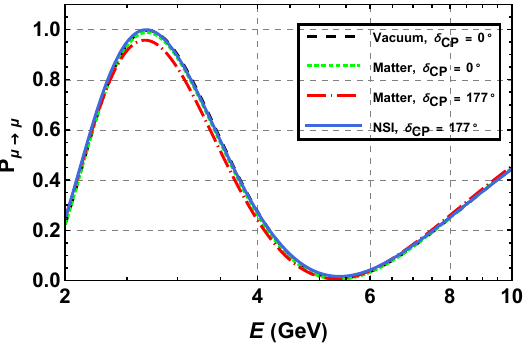}
        \caption{P2O}
        \label{fig:pe1k}
    \end{subfigure}
    \hfill
    \begin{subfigure}[b]{0.33\textwidth}
        \centering
        \includegraphics[width=\textwidth]{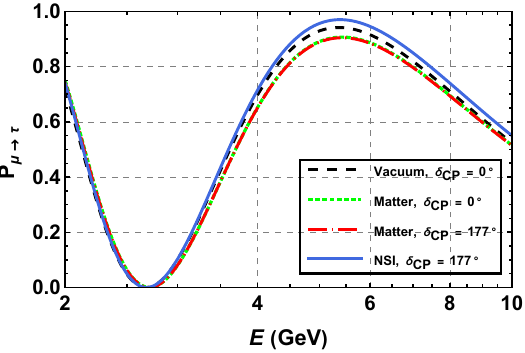}
        \caption{P2O}
        \label{fig:pe1l}
    \end{subfigure}
    \caption{In the NO scenario, we depict the transition probabilities $P_{\mu\rightarrow e}$, $P_{\mu\rightarrow \mu}$, and $P_{\mu\rightarrow \tau}$ as functions of energy $E\,(\text{GeV})$ for the initial muon flavor neutrino state $\ket{\nu_\mu}$, considering evolution in vacuum, matter, and in the presence of NSI parameters, with two distinct values of $\delta_{\text{CP}}$. Their comparisons are illustrated by fixing the baseline lengths corresponding to the T2K, NOvA, DUNE, and P2O experiments in Figs.\,(\ref{fig:pe1a}, \ref{fig:pe1b}, \ref{fig:pe1c}), Figs.\,(\ref{fig:pe1d}, \ref{fig:pe1e}, \ref{fig:pe1f}),
Figs.\,(\ref{fig:pe1g}, \ref{fig:pe1h}, \ref{fig:pe1i}), and Figs.\,(\ref{fig:pe1j}, \ref{fig:pe1k}, \ref{fig:pe1l}), respectively. The NSI parameters, the fundamental vacuum parameters for NO, and the baseline lengths for different neutrino experiments used in our analysis are taken from Tables\,\ref{tab:1}, \ref{tab:2}, and \ref{tab:3}, respectively.}
    \label{fig:prob-vs-energy-1}
\end{figure}

\begin{figure}[H]
    \centering
    \begin{subfigure}[b]{0.33\textwidth}
        \centering
        \includegraphics[width=\textwidth]{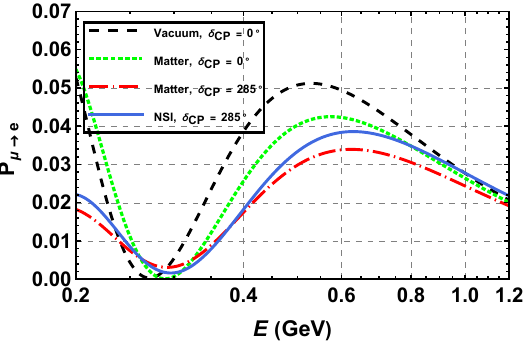}
        \caption{T2K}
        \label{fig:pe2a}
    \end{subfigure}
    \hfill
    \begin{subfigure}[b]{0.33\textwidth}
        \centering
        \includegraphics[width=\textwidth]{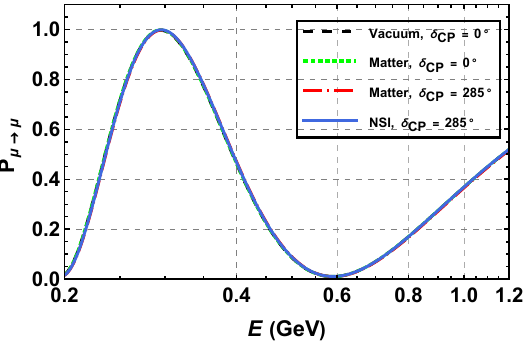}       \caption{T2K}
        \label{fig:pe2b}
    \end{subfigure}
    \hfill
    \begin{subfigure}[b]{0.33\textwidth}
        \centering
        \includegraphics[width=\textwidth]{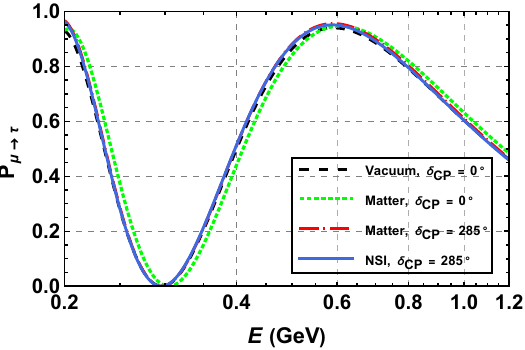}
        \caption{T2K}
        \label{fig:pe2c}
    \end{subfigure}
    \begin{subfigure}[b]{0.33\textwidth}
        \centering
        \includegraphics[width=\textwidth]{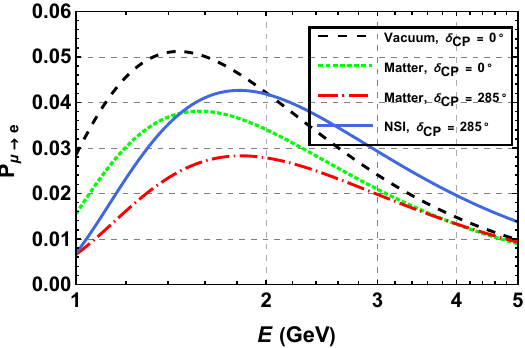}
        \caption{NOvA}
        \label{fig:pe2d}
    \end{subfigure}
    \hfill
    \begin{subfigure}[b]{0.33\textwidth}
        \centering
        \includegraphics[width=\textwidth]{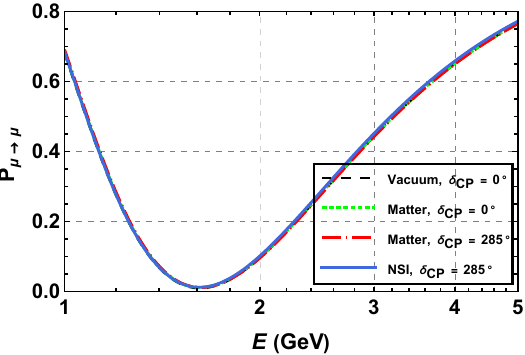}
        \caption{NOvA}
        \label{fig:pe2e}
    \end{subfigure}
    \hfill
    \begin{subfigure}[b]{0.33\textwidth}
        \centering
        \includegraphics[width=\textwidth]{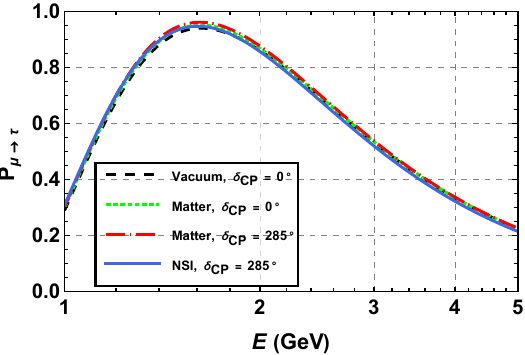}
        \caption{NOvA}
        \label{fig:pe2f}
    \end{subfigure}
    \begin{subfigure}[b]{0.33\textwidth}
        \centering
        \includegraphics[width=\textwidth]{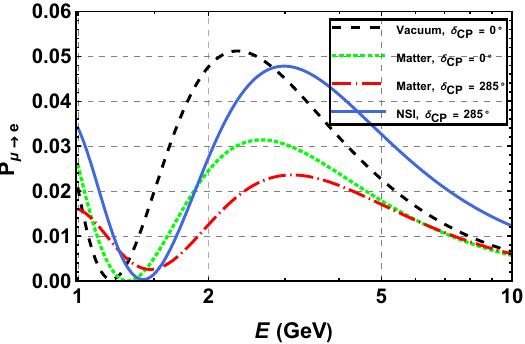}
        \caption{DUNE}
        \label{fig:pe2g}
    \end{subfigure}
    \hfill
    \begin{subfigure}[b]{0.33\textwidth}
        \centering
        \includegraphics[width=\textwidth]{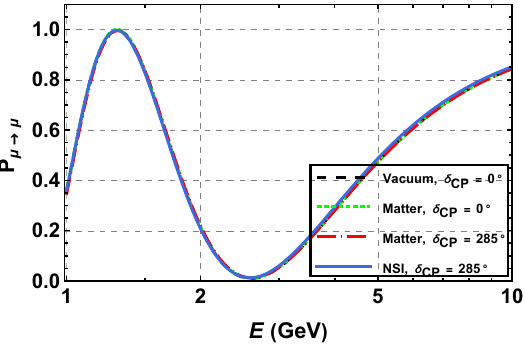}
        \caption{DUNE}
        \label{fig:pe2h}
    \end{subfigure}
    \hfill
    \begin{subfigure}[b]{0.3\textwidth}
        \centering
        \includegraphics[width=\textwidth]{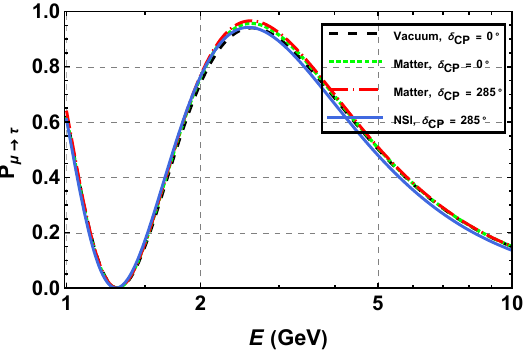}
        \caption{DUNE}
        \label{fig:pe2i}
    \end{subfigure}
    \begin{subfigure}[b]{0.33\textwidth}
        \centering
        \includegraphics[width=\textwidth]{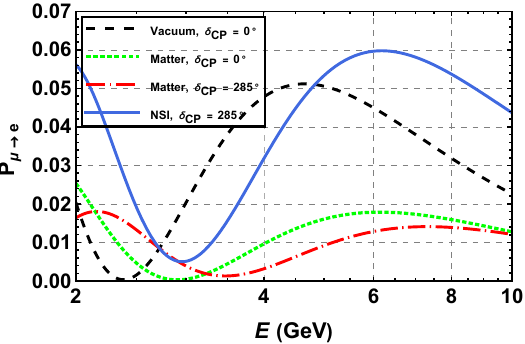}
        \caption{P2O}
        \label{fig:pe2j}
    \end{subfigure}
    \hfill
    \begin{subfigure}[b]{0.33\textwidth}
        \centering
        \includegraphics[width=\textwidth]{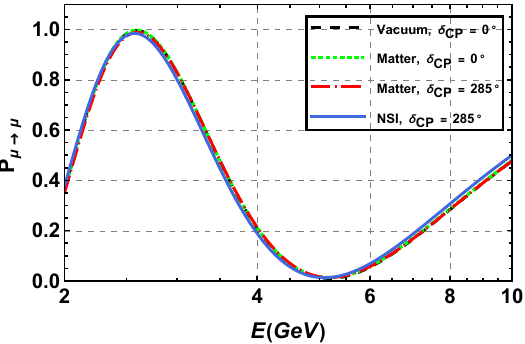}
        \caption{P2O}
        \label{fig:pe2k}
    \end{subfigure}
    \hfill
    \begin{subfigure}[b]{0.33\textwidth}
        \centering
        \includegraphics[width=\textwidth]{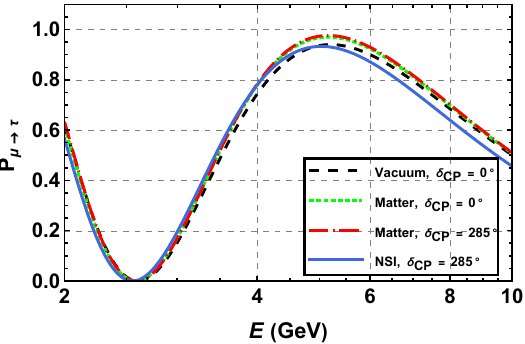}
        \caption{P2O}
        \label{fig:pe2l}
    \end{subfigure}
    \caption{In the IO scenario, we depict the transition probabilities $P_{\mu\rightarrow e}$, $P_{\mu\rightarrow \mu}$, and $P_{\mu\rightarrow \tau}$ as functions of energy $E\,(\text{GeV})$ for the initial muon flavor neutrino state $\ket{\nu_\mu}$, considering evolution in vacuum, matter, and in the presence of NSI parameters, with two distinct values of $\delta_{\text{CP}}$. Their comparisons are illustrated by fixing the baseline lengths corresponding to the T2K, NOvA, DUNE, and P2O experiments in Figs.\,(\ref{fig:pe2a}, \ref{fig:pe2b}, \ref{fig:pe2c}), Figs.\,(\ref{fig:pe2d}, \ref{fig:pe2e}, \ref{fig:pe2f}),
Figs.\,(\ref{fig:pe2g}, \ref{fig:pe2h}, \ref{fig:pe2i}), and Figs.\,(\ref{fig:pe2j}, \ref{fig:pe2k}, \ref{fig:pe2l}), respectively. The NSI parameters, the fundamental vacuum parameters for IO, and the baseline lengths for different neutrino experiments used in our analysis are taken from Tables\,\ref{tab:1}, \ref{tab:2}, and \ref{tab:3}, respectively.}
    \label{fig:prob-vs-energy-2}
\end{figure}

\begin{figure}[H]
    \centering
    \begin{subfigure}[b]{0.33\textwidth}
        \centering
        \includegraphics[width=\textwidth]{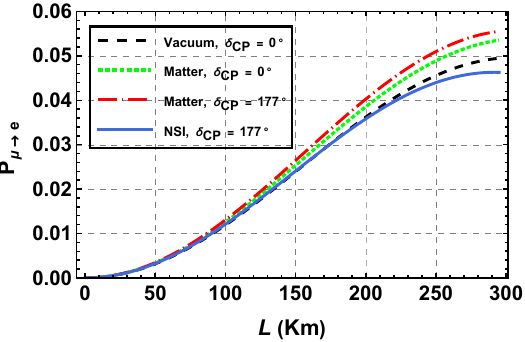}
        \caption{T2K}
        \label{fig:pe3a}
    \end{subfigure}
    \hfill
    \begin{subfigure}[b]{0.33\textwidth}
        \centering
        \includegraphics[width=\textwidth]{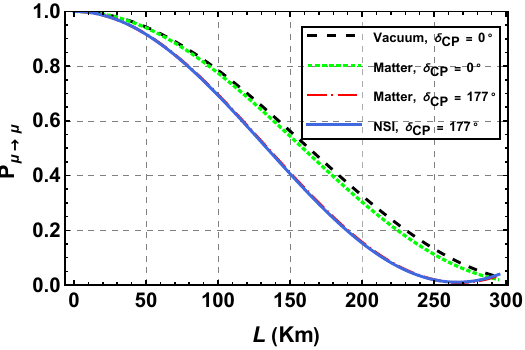}
        \caption{T2K}
        \label{fig:pe3b}
    \end{subfigure}
    \hfill
    \begin{subfigure}[b]{0.33\textwidth}
        \centering
        \includegraphics[width=\textwidth]{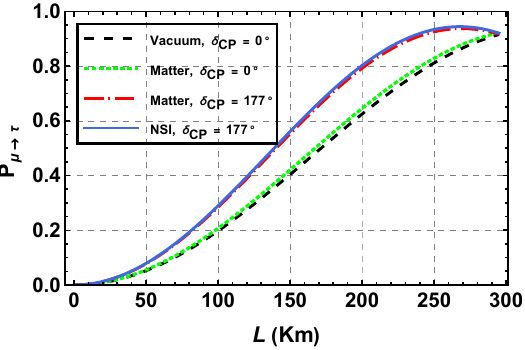}
        \caption{T2K}
        \label{fig:pe3c}
    \end{subfigure}
     \begin{subfigure}[b]{0.33\textwidth}
        \centering
        \includegraphics[width=\textwidth]{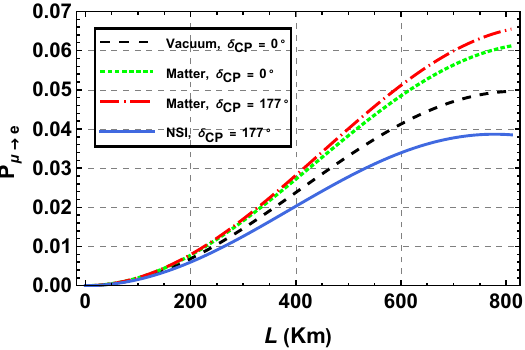}
        \caption{NOvA}
        \label{fig:pe3d}
    \end{subfigure}
    \hfill
    \begin{subfigure}[b]{0.33\textwidth}
        \centering
        \includegraphics[width=\textwidth]{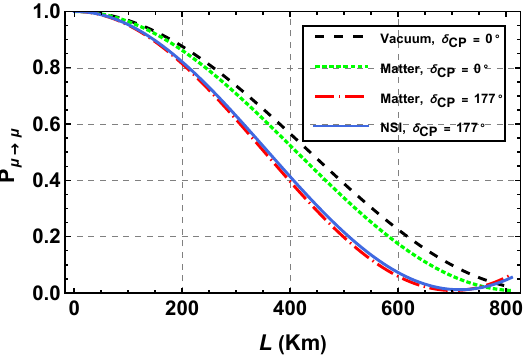}
        \caption{NOvA}
        \label{fig:pe3e}
    \end{subfigure}
    \hfill
    \begin{subfigure}[b]{0.33\textwidth}
        \centering
        \includegraphics[width=\textwidth]{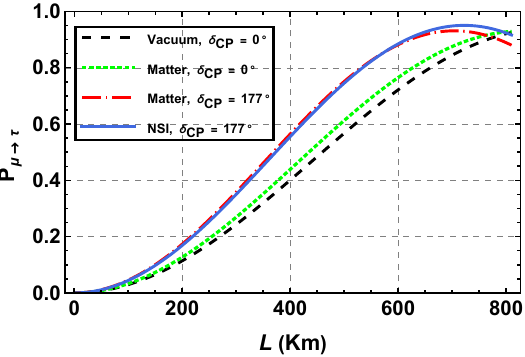}
        \caption{NOvA}
        \label{fig:pe3f}
    \end{subfigure}
    \begin{subfigure}[b]{0.3\textwidth}
        \centering
        \includegraphics[width=\textwidth]{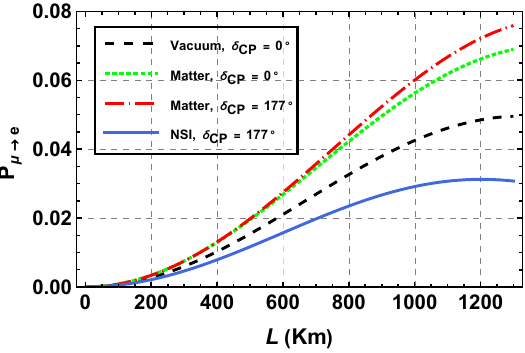}
        \caption{DUNE}
        \label{fig:pe3g}
    \end{subfigure}
    \hfill
    \begin{subfigure}[b]{0.33\textwidth}
        \centering
        \includegraphics[width=\textwidth]{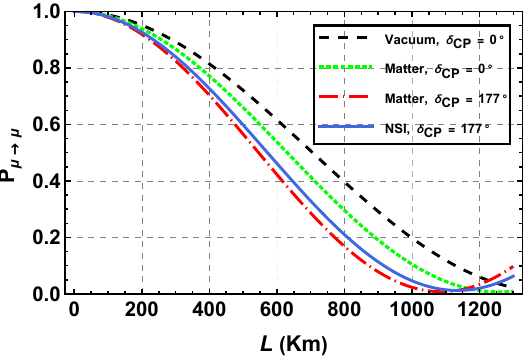}
        \caption{DUNE}
        \label{fig:pe3h}
    \end{subfigure}
    \hfill
    \begin{subfigure}[b]{0.33\textwidth}
        \centering
        \includegraphics[width=\textwidth]{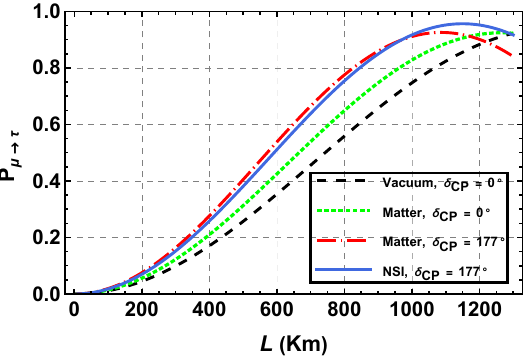}
        \caption{DUNE}
        \label{fig:pe3i}
    \end{subfigure}
    \begin{subfigure}[b]{0.33\textwidth}
        \centering
        \includegraphics[width=\textwidth]{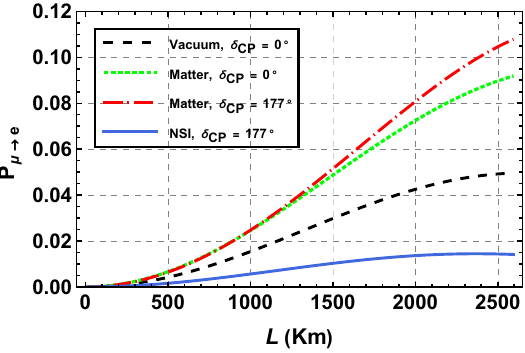}
        \caption{P2O}
        \label{fig:pe3j}
    \end{subfigure}
    \hfill
    \begin{subfigure}[b]{0.3\textwidth}
        \centering
        \includegraphics[width=\textwidth]{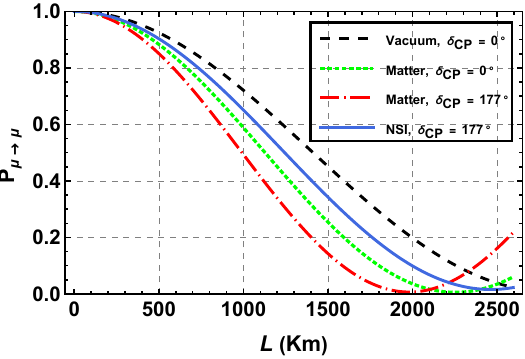}
        \caption{P2O}
        \label{fig:pe3k}
    \end{subfigure}
    \hfill
    \begin{subfigure}[b]{0.33\textwidth}
        \centering
        \includegraphics[width=\textwidth]{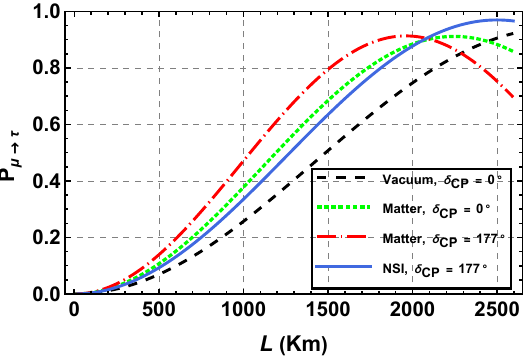}
        \caption{P2O}
        \label{fig:pe3l}
    \end{subfigure}
    \caption{In the NO scenario, we depict the transition probabilities $P_{\mu\rightarrow e}$, $P_{\mu\rightarrow \mu}$, and $P_{\mu\rightarrow \tau}$ as functions of propagation length $L\,(\text{km}$) for the initial muon flavor neutrino state $\ket{\nu_\mu}$, considering evolution in vacuum, matter, and in the presence of NSI parameters, with two distinct values of $\delta_{\text{CP}}$. Their comparisons are illustrated by fixing the neutrino energy corresponding to the T2K, NOvA, DUNE, and P2O experiments in Figs.\,(\ref{fig:pe3a}, \ref{fig:pe3b}, \ref{fig:pe3c}), Figs.\,(\ref{fig:pe3d}, \ref{fig:pe3e}, \ref{fig:pe3f}),
Figs.\,(\ref{fig:pe3g}, \ref{fig:pe3h}, \ref{fig:pe3i}), and Figs.\,(\ref{fig:pe3j}, \ref{fig:pe3k}, \ref{fig:pe3l}), respectively. The NSI parameters and the fundamental vacuum parameters for NO used in our analysis are taken from Tables \ref{tab:1} and \ref{tab:2}, respectively. The fixed neutrino energies used for the T2K, NOvA, DUNE, and P2O experiments in vacuum, matter, and NSI, with two distinct $\delta_{\text{CP}}$ phase values, are taken from Table\,\ref{tab:4}.}
    \label{fig:prob-vs-length-3}
\end{figure}

\begin{figure}[H]
    \centering
    \begin{subfigure}[b]{0.33\textwidth}
        \centering
        \includegraphics[width=\textwidth]{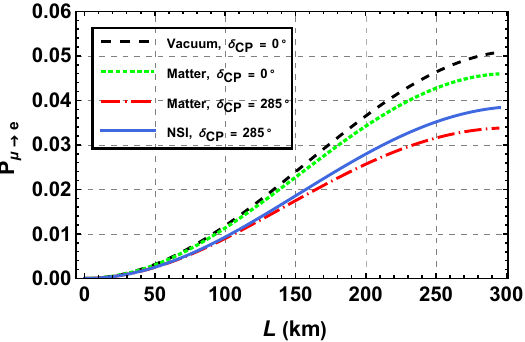}
        \caption{T2K}
        \label{fig:pe4a}
    \end{subfigure}
    \hfill
    \begin{subfigure}[b]{0.33\textwidth}
        \centering
        \includegraphics[width=\textwidth]{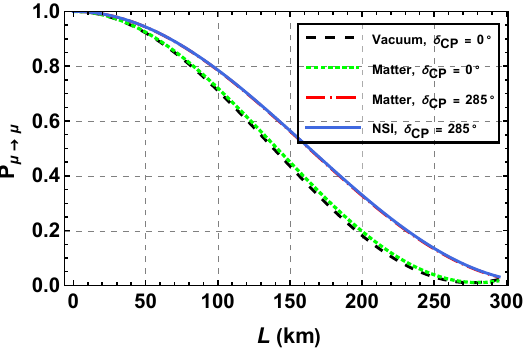}
        \caption{T2K}
        \label{fig:pe4b}
    \end{subfigure}
    \hfill
    \begin{subfigure}[b]{0.33\textwidth}
        \centering
        \includegraphics[width=\textwidth]{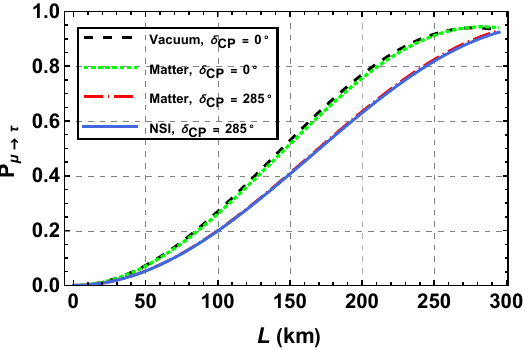}
        \caption{T2K}
        \label{fig:pe4c}
    \end{subfigure}
    \begin{subfigure}[b]{0.3\textwidth}
        \centering
        \includegraphics[width=\textwidth]{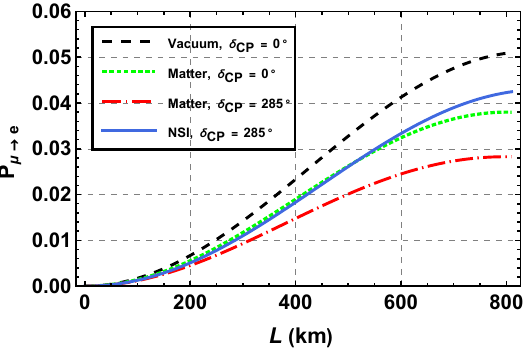}
        \caption{NOvA}
        \label{fig:pe4d}
    \end{subfigure}
    \hfill
    \begin{subfigure}[b]{0.33\textwidth}
        \centering
        \includegraphics[width=\textwidth]{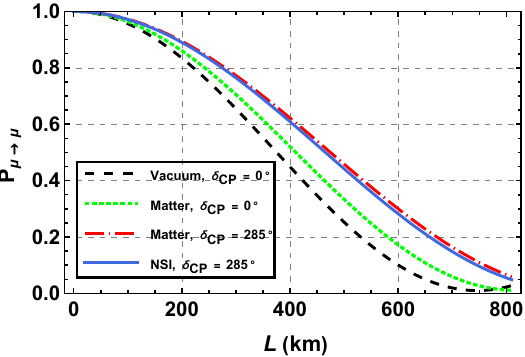}
        \caption{NOvA}
        \label{fig:pe4e}
    \end{subfigure}
    \hfill
    \begin{subfigure}[b]{0.3\textwidth}
        \centering
        \includegraphics[width=\textwidth]{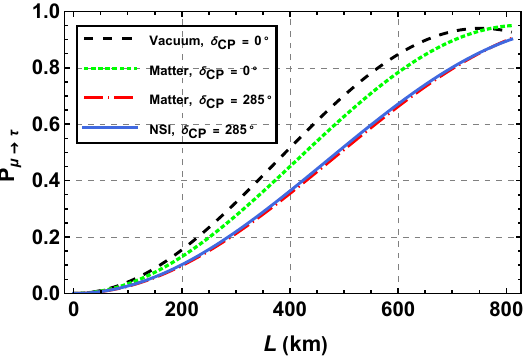}
        \caption{NOvA}
        \label{fig:pe4f}
    \end{subfigure}
    \begin{subfigure}[b]{0.33\textwidth}
        \centering
        \includegraphics[width=\textwidth]{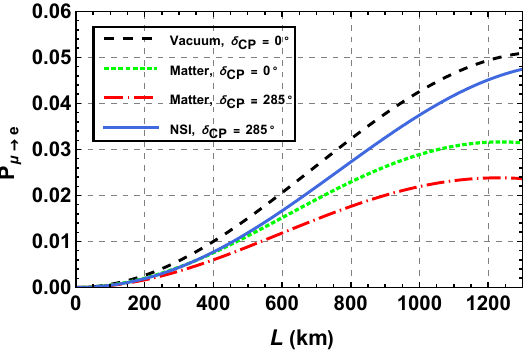}
        \caption{DUNE}
        \label{fig:pe4g}
    \end{subfigure}
    \hfill
    \begin{subfigure}[b]{0.3\textwidth}
        \centering
        \includegraphics[width=\textwidth]{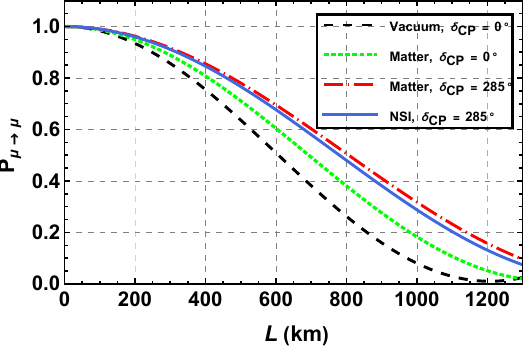}
        \caption{DUNE}
        \label{fig:pe4h}
    \end{subfigure}
    \hfill
    \begin{subfigure}[b]{0.33\textwidth}
        \centering
        \includegraphics[width=\textwidth]{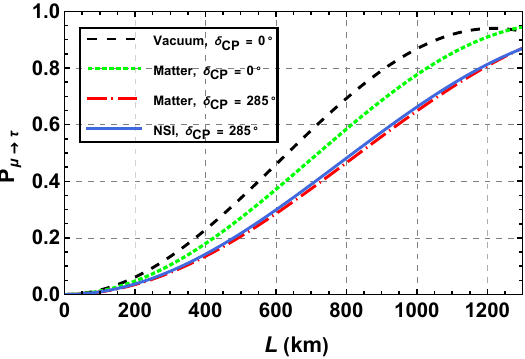}
        \caption{DUNE}
        \label{fig:pe4i}
    \end{subfigure}
    \begin{subfigure}[b]{0.33\textwidth}
        \centering
        \includegraphics[width=\textwidth]{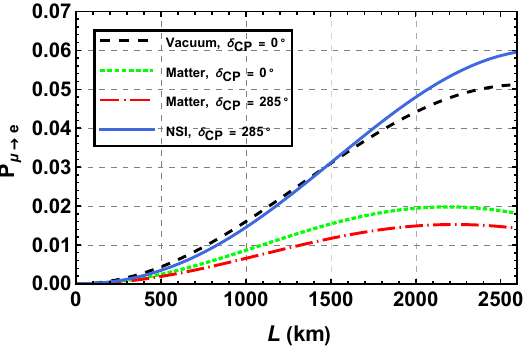}
        \caption{P2O}
        \label{fig:pe4j}
    \end{subfigure}
    \hfill
    \begin{subfigure}[b]{0.33\textwidth}
        \centering
        \includegraphics[width=\textwidth]{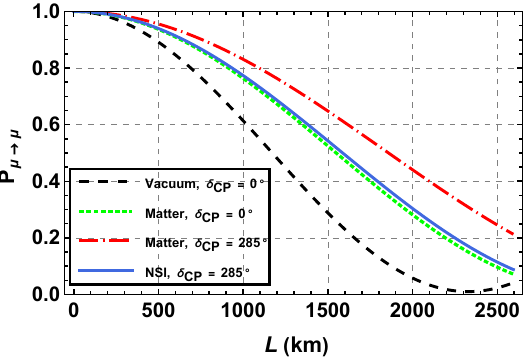}
        \caption{P2O}
        \label{fig:pe4k}
    \end{subfigure}
    \hfill
    \begin{subfigure}[b]{0.33\textwidth}
        \centering
        \includegraphics[width=\textwidth]{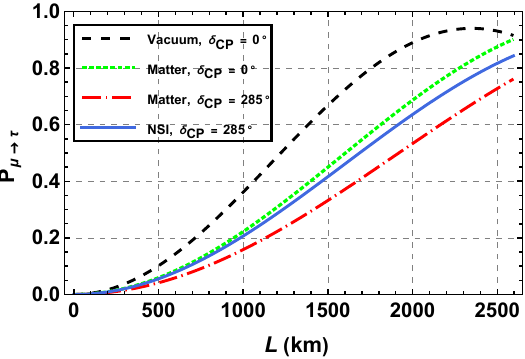}
        \caption{P2O}
        \label{fig:pe4l}
    \end{subfigure}
    \caption{In the IO scenario, we depict the transition probabilities $P_{\mu\rightarrow e}$, $P_{\mu\rightarrow \mu}$, and $P_{\mu\rightarrow \tau}$ as functions of propagation length $L\,(\text{km})$ for the initial muon flavor neutrino state $\ket{\nu_\mu}$, considering evolution in vacuum, matter, and in the presence of NSI parameters, with two distinct values of $\delta_{\text{CP}}$. Their comparisons are illustrated by fixing the neutrino energy corresponding to the T2K, NOvA, DUNE, and P2O experiments in Figs.\,(\ref{fig:pe4a}, \ref{fig:pe4b}, \ref{fig:pe4c}), Figs.\,(\ref{fig:pe4d}, \ref{fig:pe4e}, \ref{fig:pe4f}),
Figs.\,(\ref{fig:pe4g}, \ref{fig:pe4h}, \ref{fig:pe4i}), and Figs.\,(\ref{fig:pe4j}, \ref{fig:pe4k}, \ref{fig:pe4l}), respectively.  The NSI parameters and the fundamental vacuum parameters for NO used in our analysis are taken from Tables \ref{tab:1} and \ref{tab:2}, respectively. The fixed neutrino energies used for the T2K, NOvA, DUNE, and P2O experiments in vacuum, matter, and NSI, with two distinct $\delta_{\text{CP}}$ phase values, are taken from Table\,\ref{tab:5}.}
    \label{fig:prob-vs-length-4}
\end{figure}

\begin{figure}[h!]
    \centering
    \begin{subfigure}[b]{0.3\textwidth}
        \centering
        \includegraphics[width=\textwidth]{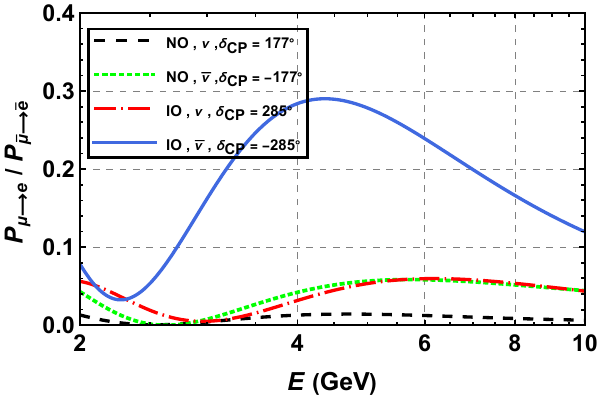 }
        \caption{P2O }
        \label{fig:pe5Ea}
    \end{subfigure}
    \hfill
    \begin{subfigure}[b]{0.3\textwidth}
        \centering
        \includegraphics[width=\textwidth]{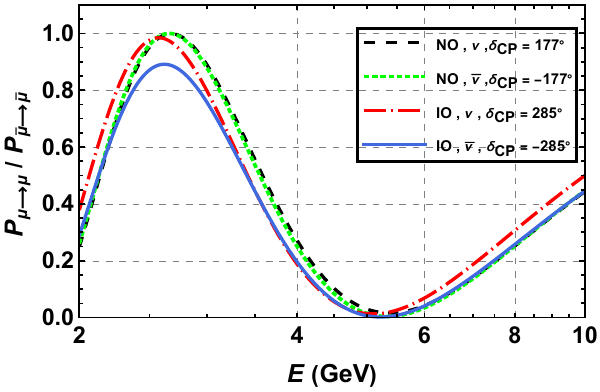}
        \caption{P2O}
        \label{fig:pe5Eb}
    \end{subfigure}
    \hfill
    \begin{subfigure}[b]{0.3\textwidth}
        \centering
        \includegraphics[width=\textwidth]{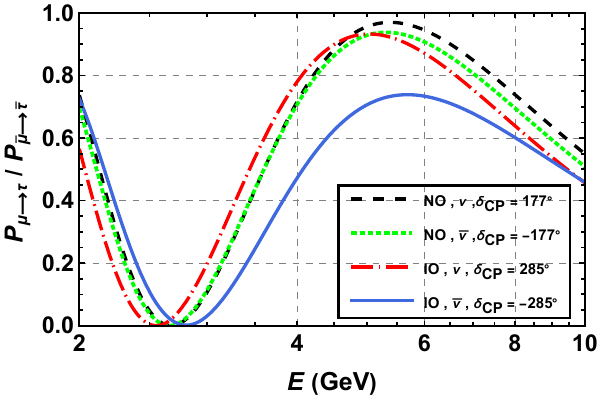}

        \caption{P2O}
        \label{fig:pe5Ec}
    \end{subfigure}
  \caption{\justifying{We depict the transition probabilities $P_{\mu\rightarrow e}$, $P_{\mu\rightarrow \mu}$, and $P_{\mu\rightarrow \tau}$ for the initial muon neutrino flavor state $\ket{\nu_\mu}$ (with positive matter potential $+V_{\rm CC}$ and positive $CP$-violation phase $+\delta_{\text{CP}}$) and  $P_{\overline{\mu} \rightarrow \overline{e}}$, $P_{\overline{\mu} \rightarrow \overline{\mu}}$, and $P_{\overline{\mu} \rightarrow \overline{\tau}}$ for the muon antineutrino flavor state $\ket{\overline{\nu}_\mu}$ (with negative matter potential $-V_{\rm CC}$ and negative $CP$-violation phase $-\delta_{\text{CP}}$)  as functions of energy $E\,(\text{GeV})$ in the presence of NSI parameters, incorporating both the fundamental vacuum parameters and the matter potential. These transition probabilities are compared between the NO and IO, using their respective best-fit values of $\delta_{\text{CP}}$, while fixing the baseline length of the P2O experiment, as shown in Figs.\,\ref{fig:pe5Ea}, \ref{fig:pe5Eb}, and \ref{fig:pe5Ec}, respectively. The NSI parameters, the fundamental vacuum parameters for the NO and IO scenarios, and the baseline length for the P2O neutrino experiment used in our analysis are taken from Tables\,\ref{tab:1}, \ref{tab:2}, and \ref{tab:3}, respectively.}}
    \label{fig:prob-vs-energy-neutrino-antineutrino-P2O-5E}
\end{figure}


\begin{figure}[H]
    \centering
    \begin{subfigure}[b]{0.33\textwidth}
        \centering
        \includegraphics[width=\textwidth]{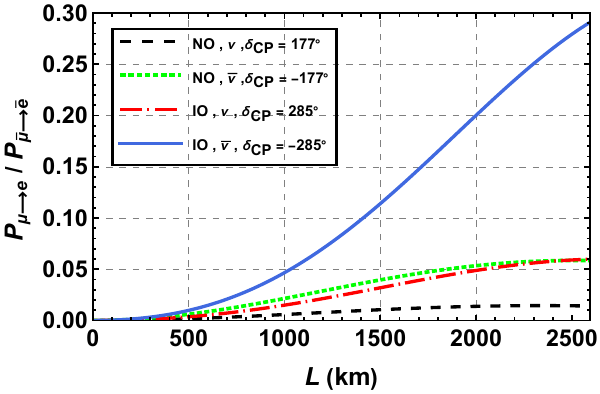}
        \caption{P2O}
        \label{fig:pe5La}
    \end{subfigure}
    \hfill
    \begin{subfigure}[b]{0.33\textwidth}
        \centering
        \includegraphics[width=\textwidth]{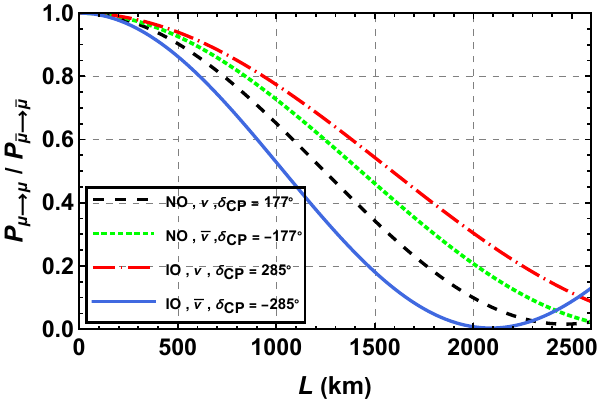}
        \caption{P2O}
        \label{fig:pe5Lb}
    \end{subfigure}
    \hfill
    \begin{subfigure}[b]{0.33\textwidth}
        \centering
        \includegraphics[width=\textwidth]{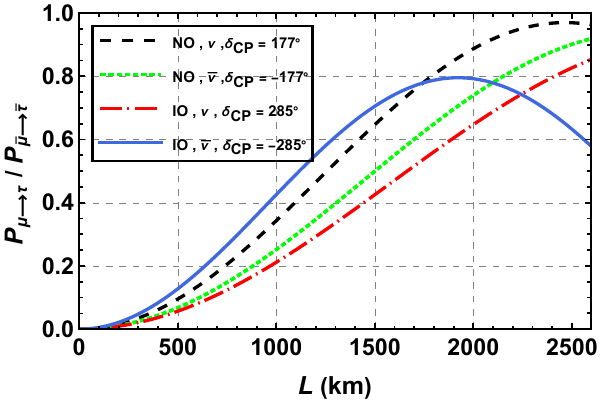}
        \caption{P2O}
        \label{fig:pe5Lc}
    \end{subfigure}
    \caption{We depict the transition probabilities $P_{\mu\rightarrow e}$, $P_{\mu\rightarrow \mu}$, and $P_{\mu\rightarrow \tau}$ for the initial muon neutrino flavor state $\ket{\nu_\mu}$ (with positive matter potential $+V_{\rm CC}$ and positive $CP$-violation phase $+\delta_{\text{CP}}$) and  $P_{\overline{\mu} \rightarrow \overline{e}}$, $P_{\overline{\mu} \rightarrow \overline{\mu}}$, and $P_{\overline{\mu} \rightarrow \overline{\tau}}$ for the muon antineutrino flavor state $\ket{\overline{\nu}_\mu}$ (with negative matter potential $-V_{\rm CC}$ and negative $CP$-violation phase $-\delta_{\text{CP}}$) as functions of propagation length $L\,(\text{km})$ in the presence of NSI parameters, incorporating both the fundamental vacuum parameters and the matter potential. These transition probabilities are compared between the NO and IO, using their respective best-fit values of $\delta_{\text{CP}}$, while fixing the neutrino energy of the P2O experiment, as shown in Figs.\,\ref{fig:pe5La}, \ref{fig:pe5Lb}, and \ref{fig:pe5Lc}, respectively. The NSI parameters, the fundamental vacuum parameters, and the fixed neutrino energy of the P2O experiment for the NO and IO scenarios, used in our analysis are taken from Tables\,\ref{tab:1}, \ref{tab:2}, and \ref{tab:6}, respectively.}
    \label{fig:prob-vs-length-neutrino-antineutrino-P2O-5L}
\end{figure}

Using Eq.\,(\ref{18}), Figs.\,\ref{fig:prob-vs-energy-1} and \ref{fig:prob-vs-energy-2} 
shows the transition probabilities $P_{\mu\rightarrow e}$, $P_{\mu\rightarrow \mu}$, and $P_{\mu\rightarrow \tau}$ as functions of energy $E\,(\text{GeV})$ for the initial muon flavor neutrino state $\ket{\nu_\mu}$ in both NO and IO scenarios, respectively, with baseline lengths $L\,(\text{km})$ fixed to those of long-baseline neutrino accelerator experiments, namely, T2K, NOvA, DUNE and P2O, as given in Table\,\ref{tab:3}. The NSI real parameters and the fundamental vacuum parameters in NO and IO scenarios used for Figs.\,\ref{fig:prob-vs-energy-1} and \ref{fig:prob-vs-energy-2} are taken from Tables\,\ref{tab:1} and \ref{tab:2}, respectively. Discrepancies in $P_{\mu\rightarrow e}$ are observed between different cases for both NO and IO scenarios, as shown in Figs.\,(\ref{fig:pe1a}, \ref{fig:pe1d}, \ref{fig:pe1g},  \ref{fig:pe1j}), and Figs.\,(\ref{fig:pe2a}, \ref{fig:pe2d}, \ref{fig:pe2g}, \ref{fig:pe2j}), respectively, using the baseline lengths of T2K, NOvA, DUNE, and P2O experiments.
In contrast, the differences for $P_{\mu\rightarrow \mu}$, and $P_{\mu\rightarrow \tau}$ between different cases are comparatively smaller in both NO and IO scenarios across all these experiments, as shown in Figs.\,(\ref{fig:pe1b}, \ref{fig:pe1c}, \ref{fig:pe1e}, \ref{fig:pe1f}, \ref{fig:pe1h}, \ref{fig:pe1i}, \ref{fig:pe1k}, \ref{fig:pe1l}), and Figs.\,(\ref{fig:pe2b}, \ref{fig:pe2c}, \ref{fig:pe2e}, \ref{fig:pe2f}, \ref{fig:pe2h}, \ref{fig:pe2i}, \ref{fig:pe2k}, \ref{fig:pe2l}), respectively.
Moreover, the energies corresponding to the first oscillation maximum peak of $P_{\mu\rightarrow e}$ for these cases in the NO and IO scenarios are shown in Figs.\,(\ref{fig:pe1a}, \ref{fig:pe1d}, \ref{fig:pe1g}, \ref{fig:pe1j}), and Figs.\,(\ref{fig:pe2a}, \ref{fig:pe2d}, \ref{fig:pe2g}, \ref{fig:pe2j}), respectively. The first oscillation maximum energy values observed for different cases are summarized in Tables\,\ref{tab:4} and \ref{tab:5} under the NO and IO scenarios, respectively.

\begin{table}[h!]
\centering
\begin{tabular}{|l|c|c|}
\hline
Mode & NO & IO \\
\hline
& $E\,(\text{GeV})$ & $E\,(\text{GeV})$\\
\hline
Muon neutrino     & 5.10  & 6.18  \\
Muon antineutrino & 5.85   & 4.20  \\
\hline
\end{tabular}
\caption{\justifying{In the NO and IO scenarios, the energy $E\,(\text{GeV})$ corresponding to the first oscillation maximum of the initial muon-flavor neutrino state $\ket{\nu_\mu}$ and muon flavor antineutrino state $\ket{\overline{\nu}_\mu}$ is determined in the presence of NSI parameters, incorporating both the fundamental vacuum parameters and a constant matter potential. These results are obtained from the transition probabilities $P_{\mu\rightarrow e}$ and $P_{\overline{\mu} \rightarrow \overline{e}}$ as shown in Fig.\,\ref{fig:pe5Ea}, for the P2O experiment.}}
\label{tab:6}
\end{table}

For the different cases in the NO and IO scenarios, Figs.\,\ref{fig:prob-vs-length-3} and \ref{fig:prob-vs-length-4}, respectively, illustrate the transition probabilities $P_{\mu\rightarrow e}$, $P_{\mu\rightarrow \mu}$, and $P_{\mu\rightarrow \tau}$ as functions of propagation length $L\,(\text{km})$ for the initial muon flavor neutrino state $\ket{\nu_\mu}$. We have used the fixed energy $E\,(\text{GeV})$ corresponding to long-baseline neutrino accelerator experiments, including T2K, NOvA, DUNE, and P2O.  The fixed energies used in our analysis for the different cases, such as vacuum, constant matter potential, and NSI, incorporating distinct values of $\delta_{\text{CP}}$ in the NO and IO scenarios, are provided in Tables\,\ref{tab:4} and \ref{tab:5}, respectively. Discrepancies in the transition probability $P_{\mu\rightarrow e}$ at the end of the baseline length across all experiments are shown in Figs.\,(\ref{fig:pe3a}, \ref{fig:pe3d}, \ref{fig:pe3g},\ref{fig:pe3j}) for NO scenario, and Figs.\,(\ref{fig:pe4a}, \ref{fig:pe4d}, \ref{fig:pe4g},\ref{fig:pe4j}) for IO scenario. These discrepancies occur among the cases of vacuum, constant matter potential, and NSI, with their respective $\delta_{\text{CP}}$ phase values. Further, it is observed that, at the end of the baseline lengths for T2K, NOvA, DUNE and P2O in the NO scenario, $P_{\mu\rightarrow e}$ is maximum for the matter potential case with $\delta_{CP}=177^\circ$ (red dot-dashed line) compared to the other cases across all these experiments, as shown in Figs.\,\ref{fig:pe3a}, \ref{fig:pe3d}, \ref{fig:pe3g}, and \ref{fig:pe3j}, respectively. However, in the IO scenario, $P_{\mu\rightarrow e}$ is maximum for vacuum with $\delta_{CP}=0^\circ$ (black dashed line) at the end of the baseline lengths of T2K, NOvA, and DUNE, as shown in Figs.\,\ref{fig:pe4a}, \ref{fig:pe4d}, and \ref{fig:pe4g}, respectively. For P2O, however, Fig.\,\ref{fig:pe4j} shows that $P_{\mu\rightarrow e}$ is maximum at the end of the baseline length in the presence of real NSI parameters with $\delta_{CP}=285^\circ$ (blue solid line). Furthermore, $P_{\mu\rightarrow \mu}$ and $P_{\mu\rightarrow \tau}$ show minimal differences among the cases of vacuum, a constant matter potential, and NSI, with their respective $\delta_{\text{CP}}$ phase values, at the end of baseline lengths for T2K and NOvA in the NO scenario, as shown in Figs.\,(\ref{fig:pe3b}, \ref{fig:pe3c}), and Figs.\,( \ref{fig:pe3e},\ref{fig:pe3f}), respectively. However, significant differences in $P_{\mu\rightarrow \mu}$ and $P_{\mu\rightarrow \tau}$ are observed at the end of baseline lengths of DUNE and P2O in NO scenario, as shown in Figs.\, (\ref{fig:pe3h}, \ref{fig:pe3i}) and Figs.\,(\ref{fig:pe3k}, \ref{fig:pe3l}), respectively. Similarly, in the IO scenarios, $P_{\mu\rightarrow \mu}$ and $P_{\mu\rightarrow \tau}$ exhibit minimal differences among the cases of vacuum, constant matter potential, and NSI, with their respective $\delta_{\text{CP}}$ phase values, at the end of the baseline lengths for T2K and NOvA, as shown in Figs.\,(\ref{fig:pe4b}, \ref{fig:pe4c}), and Figs.\,(\ref{fig:pe4e}, \ref{fig:pe4f}), respectively. In contrast, significant differences are observed among the same cases at the end of the baseline lengths for DUNE and P2O, as illustrated in Figs.\,(\ref{fig:pe4h}, \ref{fig:pe4i}), and Figs.\,(\ref{fig:pe4k}, \ref{fig:pe4l}), respectively. These results primarily arise from using the energy corresponding to the first oscillation maximum in vacuum and in a constant matter potential with NSI effects, under both the NO and IO scenarios with their respective $\delta_{\text{CP}}$ phase values.

The $CP$-violation phase $\delta_{\rm CP}$ is a key parameter influencing neutrino oscillations; however, its value is not yet well constrained in neutrino experiments \cite{T2K:2019bcf}. $\delta_{\rm CP}$ is thought to play a crucial role in explaining the matter-antimatter asymmetry of the universe, where baryons outnumber antibaryons \cite{Canetti:2012zc}. Although direct evidence for baryon number violation has not yet been observed, reactions that create an imbalance in lepton number, where leptons produced differ from those destroyed, could indirectly induce changes in baryon number~\cite{Abada:2006ea,Burguet-Castell:2001ppm,Buchmuller:2003gz,Gribov:1968kq,Davidson:2008bu,Buchmuller:2004nz,Buchmuller:2005eh,Akhmedov:1998qx,Nardi:2006fx,Frampton:2002qc,Buchmuller:2002rq,Pascoli:2006ci,Branco:2011zb}. Neutrinos, as leptons, provide a promising avenue to explore this connection. By analyzing differences in the oscillation behavior of neutrinos and antineutrinos, current long-baseline accelerator neutrino experiments such as T2K and NOvA ~\cite{DeRujula:1998umv,T2K:2019bcf,Giganti:2017fhf,T2K:2019bcf,PhysRevLett.107.041801,ABE2011106,Wachala:2017pey,NOvA:2024lti} and the upcoming DUNE~\cite{DUNE:2020ypp,Brahma:2023pxj,LBNE:2013dhi} and P2O \cite{Akindinov_2019_LetterOfInterest_P2O,Panda:2024ioo,Singha:2022btw,Majhi:2022fed,Kaur:2021rau,Parveen:2023ixk} experiment aim to detect possible signatures of $CP$ violation.

So far, we have considered the transition probabilities for the initial muon-flavor neutrino state $\ket{\nu_\mu}$ for the different cases in NO and IO scenarios.  For completeness, we now examine the transition probabilities for the initial muon-flavor antineutrino state $\ket{\overline{\nu}_\mu}$ under the same conditions, namely, in the NO and IO scenarios, with real NSI parameters, including both the fundamental vacuum parameters and a constant matter potential, and with the corresponding best-fit $\delta_{\rm CP}$ phase values.

Using Eq.\,(\ref{18}), Fig.\,\ref{fig:prob-vs-energy-neutrino-antineutrino-P2O-5E} shows the transition probabilities $P_{\mu\rightarrow e}$, $P_{\mu\rightarrow \mu}$, and $P_{\mu\rightarrow \tau}$ as functions of energy E(\text{GeV})  for the initial muon flavor neutrino state $\ket{\nu_\mu}$ and  $P_{\overline{\mu} \rightarrow \overline{e}}$, $P_{\overline{\mu} \rightarrow \overline{\mu}}$, and $P_{\overline{\mu} \rightarrow \overline{\tau}}$ for the initial muon flavor antineutrino state $\ket{\overline{\nu}_\mu}$, in both the NO and IO scenarios. For the initial state $\ket{\nu_\mu}$ case, a positive matter potential  $+V_{\rm CC}$ is used, along with a best-fit $CP$-violation phase of $\delta_{\text{CP}}=177^{\circ}$ for NO and  $\delta_{\text{CP}}=285^\circ$ for IO. In contrast, for the initial state  $\ket{\overline{\nu}_\mu}$, a negative matter potential  $-V_{\rm CC}$ is used, with corresponding best-fit $CP$-violation phases of $\delta_{\text{CP}}=-177^\circ$ for NO and  $\delta_{\text{CP}}=-285^\circ$ for IO. These transition probabilities are evaluated at the fixed baseline length ($L\approx 2595\,\rm{km}$) of the P2O experiment, under the influence of real NSI parameters together with both the fundamental vacuum parameters and a constant matter potential. As shown in Fig.\,\ref{fig:pe5Ea}, a comparison between the NO and IO scenarios reveals significant discrepancies in $P_{\mu\rightarrow e}$ and $P_{\overline{\mu} \rightarrow \overline{e}}$ in the presence of real NSI parameters, using the respective best-fit $\delta_{\rm CP}$ values. For the initial state $\ket{\overline{\nu}_\mu}$ in IO scenario, the results indicate that $P_{\overline{\mu} \rightarrow \overline{e}}$ is maximum at $\delta_{\text{CP}}=-285^\circ$ (blue solid line).
Similarly, discrepancies in $P_{\mu\rightarrow \mu}$ /$P_{\overline{\mu} \rightarrow \overline{\mu}}$ and $P_{\mu\rightarrow \tau}$ / $P_{\overline{\mu} \rightarrow \overline{\tau}}$ are also observed between NO and IO in the presence of real NSI parameters, as shown in Figs.\,\ref{fig:pe5Eb} and \ref{fig:pe5Ec}, respectively. Moreover, using the fixed baseline length of the P2O experiment, in the presence of NSI parameters, the energies corresponding to the first oscillation maximum of $P_{\mu\rightarrow e}$ and $P_{\overline{\mu} \rightarrow \overline{e}}$ in both NO and IO scenarios, incorporating their respective best-fit $\delta_{\rm CP}$ values for the initial states $\ket{\nu_\mu}$ and $\ket{\overline{\nu}_\mu}$ are identified from Fig.\,\ref{fig:pe5Ea}, and are summarized in Table\,\ref{tab:6}.

In Fig.\,\ref{fig:prob-vs-length-neutrino-antineutrino-P2O-5L}, we use the fixed energies from Table\,\ref{tab:6} to describe the transition probabilities $P_{\mu\rightarrow e}$, $P_{\mu\rightarrow \mu}$, and $P_{\mu\rightarrow \tau}$ as functions of the propagation length $L(\text{GeV})$ for the initial muon flavor neutrino state $\ket{\nu_\mu}$, and  $P_{\overline{\mu} \rightarrow \overline{e}}$, $P_{\overline{\mu} \rightarrow \overline{\mu}}$, and $P_{\overline{\mu} \rightarrow \overline{\tau}}$ for the initial muon flavor antineutrino state $\ket{\overline{\nu}_\mu}$, under both the NO and IO scenarios. These transition probabilities are evaluated in the presence of real NSI parameters, incorporating both the fundamental vacuum parameters and a constant matter, with their respective best-fit $\delta_{\text{CP}}$ phase values, and compared between the NO and IO scenarios. Figures\,\ref{fig:pe5La}, \ref{fig:pe5Lb}, and \ref{fig:pe5Lc} show discrepancies in $P_{\mu\rightarrow e}$ / $P_{\overline{\mu} \rightarrow \overline{e}}$, $P_{\mu\rightarrow \mu}$ / $P_{\overline{\mu} \rightarrow \overline{\mu}}$, and  $P_{\mu\rightarrow \tau}$ / $P_{\overline{\mu} \rightarrow \overline{\tau}}$, respectively, for the initial states $\ket{\nu_\mu}$ / $\ket{\overline{\nu}_\mu}$, between NO and IO with their best-fit $\delta_{\text{CP}}$ phase values.

Building on the transition probabilities in the three-flavor scenario, we now take up the case of the quantum spread complexity in three-flavor neutrino oscillations.

\section{Quantum Spread Complexity in Neutrino Oscillation with NSI}
\label{Sec4}
 In this section, we investigate the quantum spread complexity as a primary analytical tool to describe neutrino oscillations in the presence of NSI, which is linked to the neutrino flavor transition probabilities. In the case of two-flavor neutrino oscillations in vacuum, the cost function $X_{\mu}$ is equal to $P_{\mu\rightarrow e}$~\cite{Dixit:2023fke}. However, for three-flavor neutrino oscillations, the cost function analysis becomes more intricate in the presence of NSI.

In the three neutrino system, the initial flavor states are represented as follows~\cite{Dixit:2023fke}
\begin{align}
 |\nu_e\rangle &= \begin{bmatrix}
     1 \\
     0 \\
     0 
 \end{bmatrix}, &\nonumber
 |\nu_\mu\rangle &= \begin{bmatrix}
     0 \\
     1 \\
     0 
 \end{bmatrix}, &\nonumber
 |\nu_\tau\rangle &= \begin{bmatrix}
     0 \\
     0 \\
     1 
 \end{bmatrix}. 
\end{align}

For the case where the system starts as a pure muon flavor neutrino state $\ket{\nu_\mu}$, we choose the initial condition $|\nu_\mu\rangle = (0, 1, 0)^T$. The set of states, $\{|\psi_n\rangle\}$, is then constructed by repeatedly applying the effective Hamiltonian $\mathcal{H}_{\text{total}}$, as defined in Eq.\,\eqref{11}. Following Eqs.\,\eqref{2}, \eqref{3}, \eqref{3a}, and \eqref{11}, the set of states $|\psi_n\rangle$ can be obtained as  
\begin{align}\label{19}
|\psi_0\rangle &= |\nu_\mu\rangle,\nonumber \\ 
|\psi_1\rangle &= \mathcal{H}_{total}\,|\nu_\mu\rangle, \nonumber\\
|\psi_2\rangle &=  \mathcal{H}_{total}^2\,|\nu_\mu\rangle, \nonumber\\
|\psi_3\rangle &=  \mathcal{H}_{total}^3\,|\nu_\mu\rangle, \\
&\quad \vdots \nonumber
\end{align} 
 
These states are not orthonormal; therefore, we apply the Gram-Schmidt procedure to construct an orthonormal set known as the Krylov basis. Moreover, using Eq.\,\eqref{19} in Eqs.\,\eqref{4a} and \eqref{4}, the Krylov basis\footnote{In the three-flavor neutrino oscillation, there are three sets of non-zero Krylov basis $\{|K_n\rangle_{e}\}$, $\{|K_n\rangle_{\mu}\}$ and $\{|K_n\rangle_{\tau}\}$ \cite{Dixit:2023fke}. However, in this work, we explicitly explore the Krylov basis for the initial muon neutrino flavor state $|\nu_\mu\rangle$.} ${|K_{n}\rangle_{\mu}}$, corresponding to the initial muon flavor neutrino state $\ket{\nu_\mu}$ in the presence of NSI, incorporating both the fundamental vacuum parameters and a constant matter potential, with $\delta_{\text{CP}}$ phase values, is obtained as
\begin{align}\label{21a}
    |K_0\rangle_{\mu} \equiv |\nu_\mu\rangle = (0, 1, 0)^T, &\nonumber\\
    |K_1\rangle_{\mu} = N^{NSI}_{1\mu} (a_1, 0, a_2)^T, &\nonumber\\
    |K_2\rangle_{\mu} = N^{NSI}_{2\mu} (b_1, 0, b_2)^T,
\end{align}
where
\begin{equation}\label{21}
N^{NSI}_{1\mu} 
= \frac{1}{\sqrt{|a_1|^2 + |a_2|^2}},\ \ 
N^{NSI}_{2\mu} 
= \frac{1}{\sqrt{|b_1|^2 + |b_2|^2}}.
\end{equation}

The expressions for $a_1$, $a_2$, $b_1$, and $b_2$ are provided in the Appendix\,\ref{Append}. Using Eq.\,(\ref{17a}) and Eq.\,(\ref{21a}) in Eq.\,\eqref{5}, the cost function for the initial state $|\nu_{\mu}\rangle$ is obtained as
\begin{align}\label{24}
\chi_{\mu} = & \sum C_{n}~~|~{_\mu\langle K_{n}|}\nu_\mu(t)\rangle~|^{2} \nonumber\\=\,& 
|\text{A}_{\mu e}|^2 \left( (N^{NSI}_{1\mu})^2 |a_1|^2 + 2 (N^{NSI}_{2\mu})^2 |b_1|^2 \right)  + |\text{A}_{\mu \tau}|^2 \left( (N^{NSI}_{1\mu})^2 |a_2|^2 + 2 (N^{NSI}_{2\mu})^2 |b_2|^2 \right) \nonumber\\
& + 2\, \Re\left( \text{A}_{\mu e} \, a_2 \, (N^{NSI}_{1\mu})^2 \, \text{A}_{\mu \tau}^* \, a_1^* \right)  + 4\, \Re\left( \text{A}_{\mu e} \, b_2 \, (N^{NSI}_{2\mu})^2 \, \text{A}_{\mu \tau}^* \, b_1^* \right),
\end{align}


\begin{figure}[H]
    \centering
    \begin{subfigure}[b]{0.375\textwidth}
        \centering
     \includegraphics[width=\textwidth]{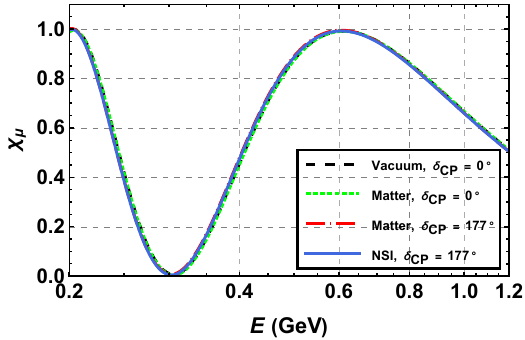}
        \caption{T2K, NO}
        \label{fig:cost-vs-energy-normal-5a}
    \end{subfigure}
    \hfill
     \begin{subfigure}[b]{0.375\textwidth}
        \centering
    \includegraphics[width=\textwidth]{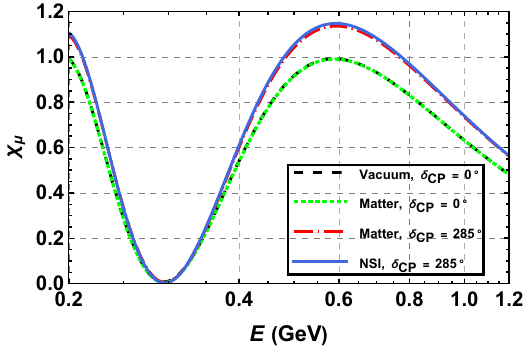}
        \caption{T2K, IO}
        \label{fig:cost-vs-energy-inverted-5b}
    \end{subfigure}
    \hfill
     \begin{subfigure}[b]{0.375\textwidth}
        \centering
    \includegraphics[width=\textwidth]{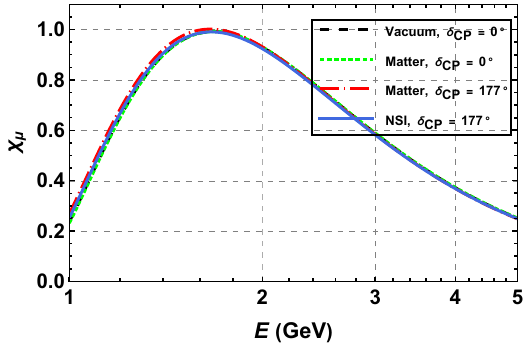}
        \caption{NOvA, NO}
        \label{fig:cost-vs-energy-normal-5c}
    \end{subfigure}
    \hfill
    \begin{subfigure}[b]{0.375\textwidth}
        \centering
    \includegraphics[width=\textwidth]{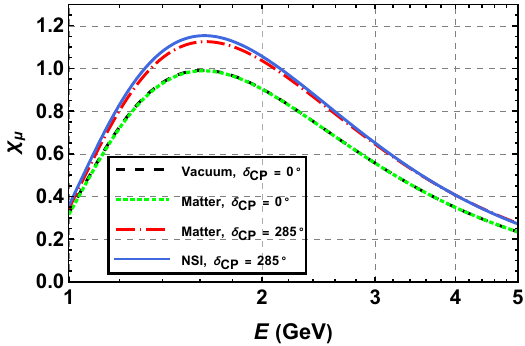}
        \caption{NOvA, IO}
        \label{fig:cost-vs-energy-inverted-5d}
    \end{subfigure}
    \hfill
     \begin{subfigure}[b]{0.375\textwidth}
        \centering
    \includegraphics[width=\textwidth]{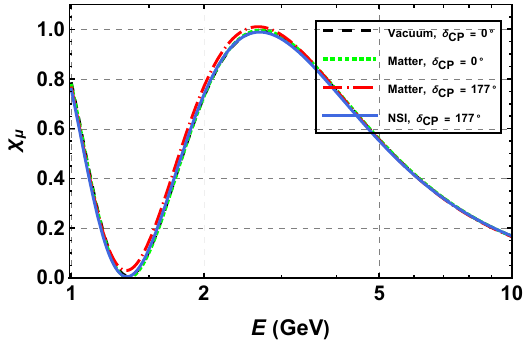}
        \caption{DUNE, NO}
        \label{fig:cost-vs-energy-normal-5e}
    \end{subfigure}
    \hfill
    \begin{subfigure}[b]{0.375\textwidth}
        \centering
   \includegraphics[width=\textwidth]{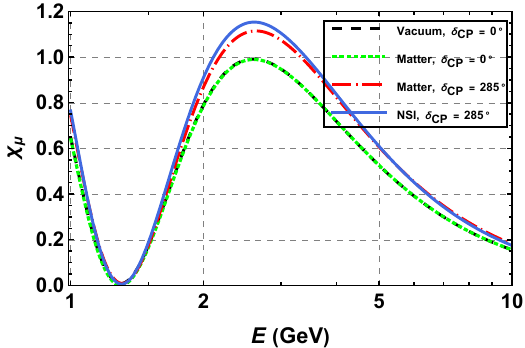}
        \caption{DUNE, IO}
        \label{fig:cost-vs-energy-inverted-5f}
    \end{subfigure}
    \hfill
    \begin{subfigure}[b]{0.375\textwidth}
        \centering
    \includegraphics[width=\textwidth]{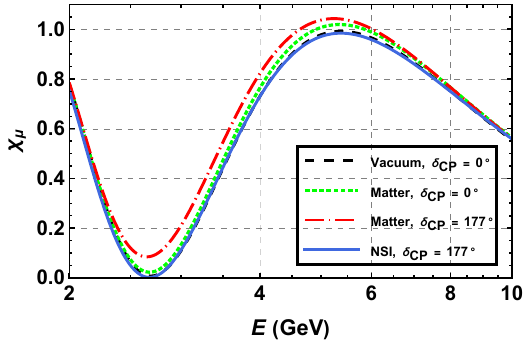}
        \caption{P2O, NO}
        \label{fig:cost-vs-energy-normal-5g}
    \end{subfigure}
    \hfill
    \begin{subfigure}[b]{0.375\textwidth}
        \centering
    \includegraphics[width=\textwidth]{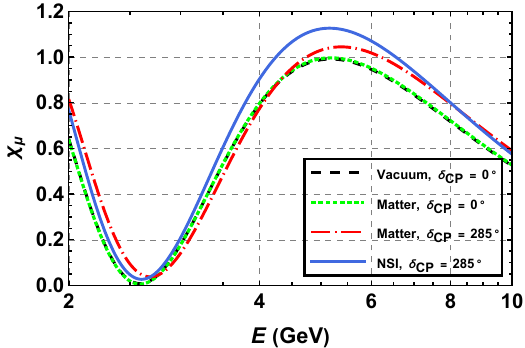}
        \caption{P2O, IO}
        \label{fig:cost-vs-energy-inverted-5h}
    \end{subfigure}
    \caption{We depict the cost function $\chi_\mu$ as functions of energy $E\,(\text{GeV})$ for the initial muon flavor neutrino state $\ket{\nu_\mu}$, considering evolution in vacuum, matter, and in the presence of NSI parameters, with two distinct values of $\delta_{\text{CP}}$. Their comparisons are illustrated in NO and IO scenarios by fixing the baseline lengths corresponding to the T2K, NOvA, DUNE, and P2O experiments in Figs.\,(\ref{fig:cost-vs-energy-normal-5a}, \ref{fig:cost-vs-energy-inverted-5b}), Figs.\,(\ref{fig:cost-vs-energy-normal-5c}, \ref{fig:cost-vs-energy-inverted-5d}),
Figs.\,(\ref{fig:cost-vs-energy-normal-5e}, \ref{fig:cost-vs-energy-inverted-5f}), and Figs.\,(\ref{fig:cost-vs-energy-normal-5g}, \ref{fig:cost-vs-energy-inverted-5h}), respectively. The NSI parameters, the fundamental vacuum parameters for the NO and IO scenarios, and the baseline lengths for different neutrino experiments used in our analysis are taken from Tables\,\ref{tab:1}, \ref{tab:2}, and \ref{tab:3}, respectively.}
    \label{fig:cost-function-vs-energy-normal-inverted-5}
\end{figure}

\begin{figure}[H]
    \centering
    \begin{subfigure}[b]{0.375\textwidth}
        \centering
     \includegraphics[width=\textwidth]{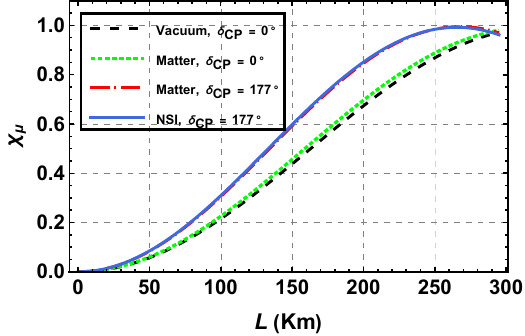}
        \caption{T2K, NO}
        \label{fig:cost-vs-length-normal-6a}
    \end{subfigure}
    \hfill
     \begin{subfigure}[b]{0.375\textwidth}
        \centering
    \includegraphics[width=\textwidth]{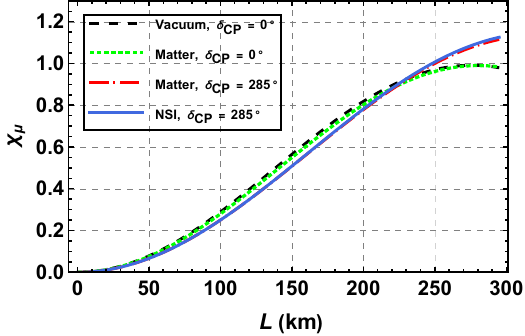}
        \caption{T2K, IO}
        \label{fig:cost-vs-length-inverted-6b}
    \end{subfigure}
    \hfill
     \begin{subfigure}[b]{0.375\textwidth}
        \centering
    \includegraphics[width=\textwidth]{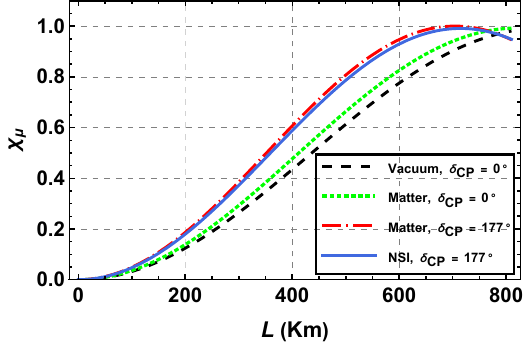}
        \caption{NOvA, NO}
        \label{fig:cost-vs-length-normal-6c}
    \end{subfigure}
    \hfill
    \begin{subfigure}[b]{0.375\textwidth}
        \centering
    \includegraphics[width=\textwidth]{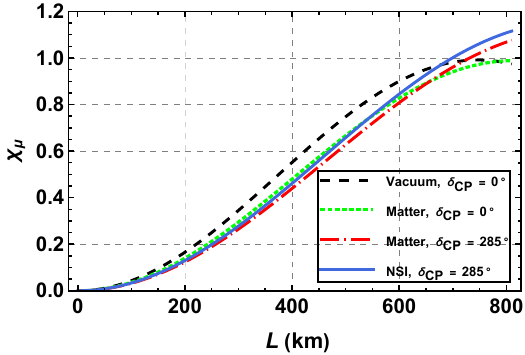}
        \caption{NOvA, IO}
        \label{fig:cost-vs-length-inverted-6d}
    \end{subfigure}
    \hfill
     \begin{subfigure}[b]{0.375\textwidth}
        \centering
    \includegraphics[width=\textwidth]{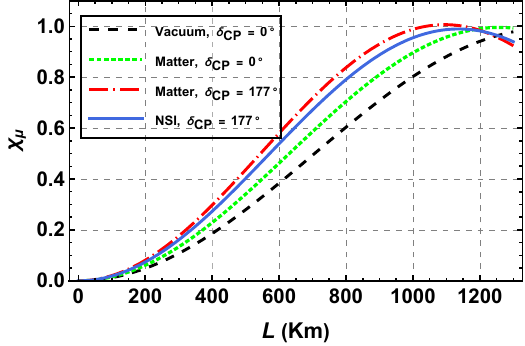}
        \caption{DUNE, NO}
        \label{fig:cost-vs-length-normal-6e}
    \end{subfigure}
    \hfill
    \begin{subfigure}[b]{0.375\textwidth}
        \centering
   \includegraphics[width=\textwidth]{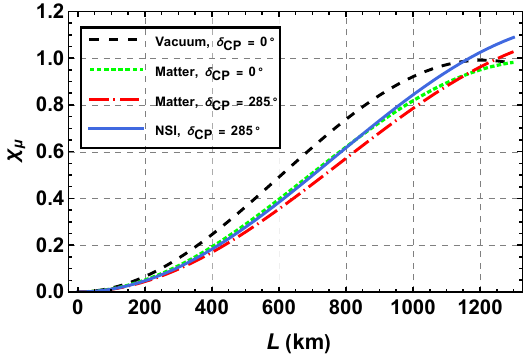}
        \caption{DUNE, IO}
        \label{fig:cost-vs-length-inverted-6f}
    \end{subfigure}
    \hfill
    \begin{subfigure}[b]{0.375\textwidth}
        \centering
    \includegraphics[width=\textwidth]{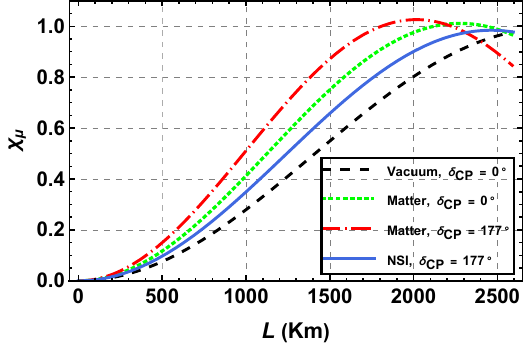}
        \caption{P2O, NO}
        \label{fig:cost-vs-length-normal-6g}
    \end{subfigure}
    \hfill
    \begin{subfigure}[b]{0.375\textwidth}
        \centering
    \includegraphics[width=\textwidth]{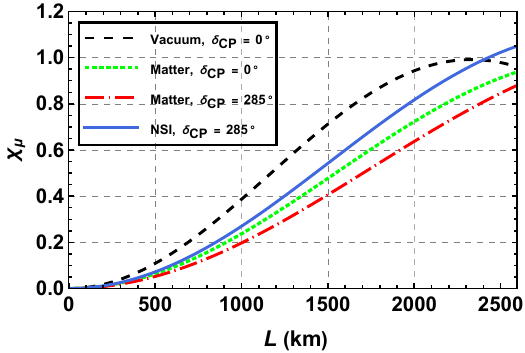}
        \caption{P2O, IO}
        \label{fig:cost-vs-length-inverted-6h}
    \end{subfigure}
    \caption{We depict the cost function $\chi_\mu$ as functions of propagation length $L\,\text{km}$) for the initial muon flavor neutrino state $\ket{\nu_\mu}$, considering evolution in vacuum, matter, and in the presence of NSI parameters, with two distinct values of $\delta_{\text{CP}}$. Their comparisons are illustrated in NO and IO scenarios by fixing the neutrino energy corresponding to the T2K, NOvA, DUNE, and P2O experiments in Figs.\,(\ref{fig:cost-vs-length-normal-6a}, \ref{fig:cost-vs-length-inverted-6b}), Figs.\,(\ref{fig:cost-vs-length-normal-6c}, \ref{fig:cost-vs-length-inverted-6d}),
Figs.\,(\ref{fig:cost-vs-length-normal-6e}, \ref{fig:cost-vs-length-inverted-6f}), and Figs.\,(\ref{fig:cost-vs-length-normal-6g}, \ref{fig:cost-vs-length-inverted-6h}), respectively. The NSI parameters, the fundamental vacuum parameters, and the fixed neutrino energies for the different experiments in the NO and IO scenarios, used in our analysis, are taken from Tables \ref{tab:1}, \ref{tab:2}, \ref{tab:4}, and \ref{tab:5}, respectively.}
    \label{fig:cost-function-vs-length-normal-inverted-6}
\end{figure}


\begin{figure}[H]
    \centering
    \begin{subfigure}[b]{0.4\textwidth}
        \centering
        \includegraphics[width=\textwidth]{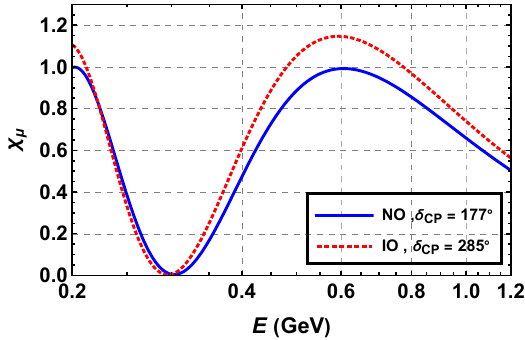}
        \caption{T2K}
        \label{fig:cost-vs-energy-comparison-NO-IO-7a}
    \end{subfigure}
    \hfill
     \begin{subfigure}[b]{0.4\textwidth}
        \centering
        \includegraphics[width=\textwidth]{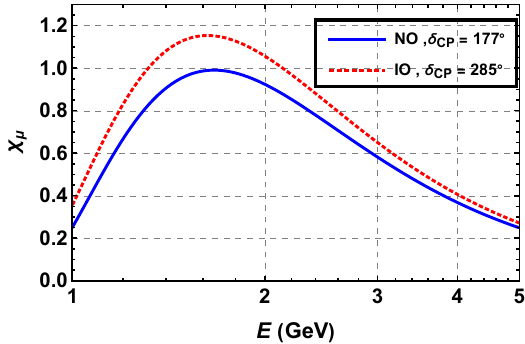}
        \caption{NOvA}
        \label{fig:cost-vs-energy-comparison-NO-IO-7b}
    \end{subfigure}
    \hfill
    \begin{subfigure}[b]{0.4\textwidth}
        \centering
        \includegraphics[width=\textwidth]{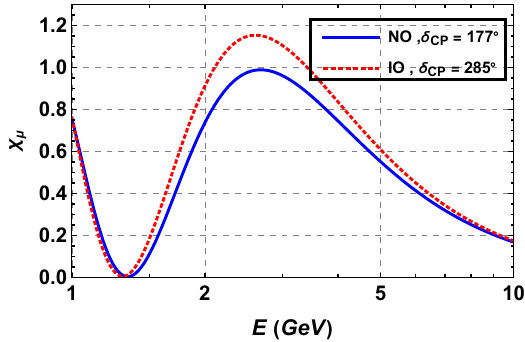}
        \caption{DUNE}
        \label{fig:cost-vs-energy-comparison-NO-IO-7c}
    \end{subfigure}
    \hfill
     \begin{subfigure}[b]{0.4\textwidth}
        \centering
        \includegraphics[width=\textwidth]{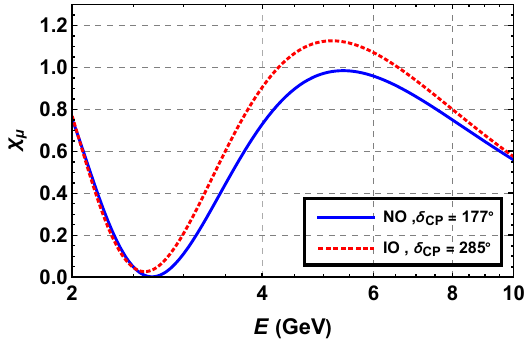}
        \caption{P2O}
        \label{fig:cost-vs-energy-comparison-NO-IO-7d}
    \end{subfigure}
    \caption{We compare the cost function $\chi_\mu$ as functions of energy $E\,(\text{GeV})$ between NO and IO scenarios for the initial muon flavor neutrino state $\ket{\nu_\mu}$, in the presence of NSI parameters. This analysis incorporates both the fundamental vacuum parameters and a constant matter potential, along with the best-fit values of $\delta_{\text{CP}}$. These comparisons are presented by fixing the baseline lengths corresponding to the T2K, NOvA, DUNE, and P2O experiments, as shown in Figs.\, \ref{fig:cost-vs-energy-comparison-NO-IO-7a}, \ref{fig:cost-vs-energy-comparison-NO-IO-7b}, \ref{fig:cost-vs-energy-comparison-NO-IO-7c}, and \ref{fig:cost-vs-energy-comparison-NO-IO-7d},  respectively. The NSI parameters, the fundamental vacuum parameters for the NO and IO scenarios, and the baseline lengths for different neutrino experiments used in our analysis are taken from Tables\,\ref{tab:1}, \ref{tab:2}, and \ref{tab:3}, respectively.}
    \label{fig:cost-function-versus-energy-NO-I0-comparison-7}
\end{figure}
where $\Re$ denotes the real part. Thus, in Eq.\,\eqref{24}, we observe that the cost function $\chi_{\mu}$ for the initial state $|\nu_\mu\rangle$ is associated with both $P_{\mu e}$ and $P_{\mu \tau}$, as well as an interference term. It is important to note that setting the NSI parameters $\epsilon_{\alpha\beta} = 0$ recovers the expression of the cost function in matter without NSI effects~\cite{Dixit:2023fke}. Furthermore, setting the constant matter potential ($V_\text{CC}$) and the $CP$-violation phase $\delta_{\text{CP}}$ to zero, we obtain the cost function corresponding to neutrino oscillations in vacuum.

Similar to the analysis of the transition probabilities in the previous section, by fixing the baseline lengths corresponding to the T2K, NOvA, DUNE, and P2O experiments in Eq.\,(\ref{24}), Fig.\,\ref{fig:cost-function-vs-energy-normal-inverted-5} compares the cost function $\chi_\mu$ as a function of energy $E\,(\text{GeV})$ for the initial muon flavor neutrino state $\ket{\nu_\mu}$. These comparisons are shown between different cases, such as the effects of NSI, as well as contributions from the fundamental vacuum parameters and a constant matter potential, using their respective $\delta_{\text{CP}}$ phase values, in both the NO and IO scenarios. We find that in the NO scenario, Figs.\,\ref{fig:cost-vs-energy-normal-5a}, \ref{fig:cost-vs-energy-normal-5c}, and \ref{fig:cost-vs-energy-normal-5e}, corresponding to the T2K, NOvA, and DUNE experiments, respectively, show minimal discrepancies in the cost function $\chi_\mu$ for the different cases. However, for the P2O experiment, the discrepancies are more pronounced among the various cases, as shown in Fig.\,\ref{fig:cost-vs-energy-normal-5g}. Furthermore, for both the DUNE and P2O experiments under the NO scenario, in Fig.\,\ref{fig:cost-vs-energy-normal-5e} and Fig.\,\ref{fig:cost-vs-energy-normal-5g}, respectively, we observe that $\chi_\mu > 1$ in the case of a constant matter potential with $\delta_{\text{CP}} = 177^\circ$ (red dot-dashed line), indicating growth in cost function during neutrino oscillations. In contrast, under the IO scenario, Figs.\,\ref{fig:cost-vs-energy-inverted-5b}, \ref{fig:cost-vs-energy-inverted-5d}, \ref{fig:cost-vs-energy-inverted-5f}, and \ref{fig:cost-vs-energy-inverted-5h}, corresponding to the T2K, NOvA, DUNE, and P2O experiments, respectively, show significant discrepancies in the cost function $\chi_\mu$ across different cases. In this scenario, $\chi_\mu > 1$ in the presence of NSI with $\delta_{\text{CP}} = 285^\circ$ (blue solid line) across all experiments, indicating that the cost function is higher compared to the other cases.

Additionally, Fig.\,\ref{fig:cost-function-vs-length-normal-inverted-6} presents the cost function $\chi_\mu$ as function of the propagation length $L\,(\text{km})$ for the initial state $\ket{\nu_\mu}$, considering various cases. The fixed energy values used in this analysis are taken from Tables\,\ref{tab:4} and \ref{tab:5}. In the NO scenario, we observe that the discrepancies in $\chi_\mu$ for different cases are relatively small at the end of the baseline lengths for the T2K and NOvA experiments, as shown in Figs.\,\ref{fig:cost-vs-length-normal-6a} and \ref{fig:cost-vs-length-normal-6c}, respectively. In contrast, the discrepancies become more pronounced at the end of the baseline lengths for the DUNE and P2O experiments, as illustrated in Figs.\,\ref{fig:cost-vs-length-normal-6e} and \ref{fig:cost-vs-length-normal-6g}. In the IO scenario, the discrepancies in $\chi_\mu$ for different cases remain small for T2K (Fig.\,\ref{fig:cost-vs-length-inverted-6b}), but are significantly larger for NOvA, DUNE, and P2O, at the end of the respective baseline length, as depicted in Figs.\,\ref{fig:cost-vs-length-inverted-6d}, \ref{fig:cost-vs-length-inverted-6f}, and \ref{fig:cost-vs-length-inverted-6h}, respectively.
\begin{figure}[H]
    \centering
    \begin{subfigure}[b]{0.34\textwidth}
        \centering
        \includegraphics[width=\textwidth]{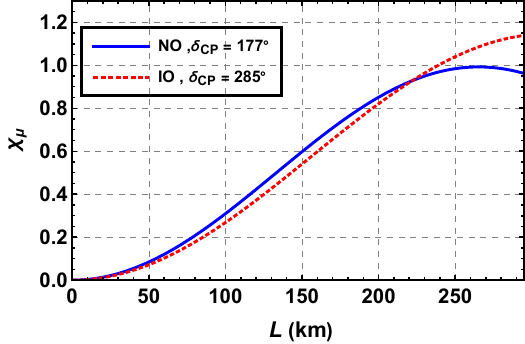}
        \caption{T2K}
        \label{fig:cost-vs-length-comparison-NO-IO-8a}
    \end{subfigure}
    \hfill
     \begin{subfigure}[b]{0.34\textwidth}
        \centering
        \includegraphics[width=\textwidth]{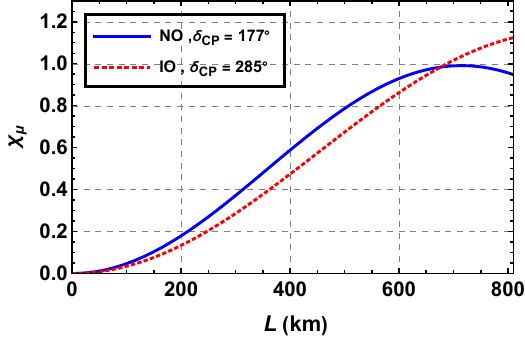}
        \caption{NOvA}
        \label{fig:cost-vs-length-comparison-NO-IO-8b}
    \end{subfigure}
    \hfill
    \begin{subfigure}[b]{0.34\textwidth}
        \centering
        \includegraphics[width=\textwidth]{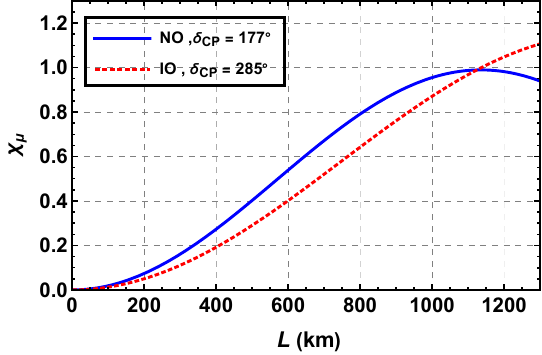}
        \caption{DUNE}
        \label{fig:cost-vs-length-comparison-NO-IO-8c}
    \end{subfigure}
    \hfill
     \begin{subfigure}[b]{0.34\textwidth}
        \centering
        \includegraphics[width=\textwidth]{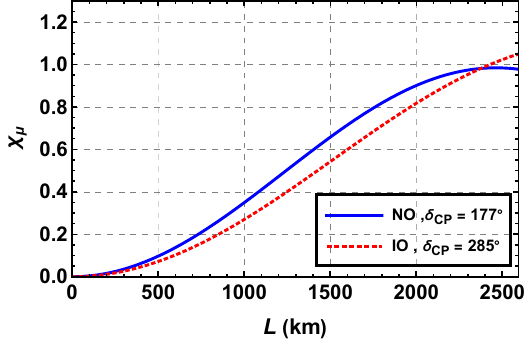}
        \caption{P2O}
        \label{fig:cost-vs-length-comparison-NO-IO-8d}
    \end{subfigure}
    \caption{We compare the cost function $\chi_\mu$ as functions of propagation length $L(\text{km})$ between NO and IO scenarios for the initial muon flavor neutrino state $\ket{\nu_\mu}$, in the presence of NSI parameters. This analysis incorporates both the fundamental vacuum parameters and a constant matter potential, along with the best-fit values of $\delta_{\text{CP}}$. These comparisons are presented by fixing the neutrino energy corresponding to the T2K, NOvA, DUNE, and P2O experiments, as shown in Figs.\,\ref{fig:cost-vs-length-comparison-NO-IO-8a}, \ref{fig:cost-vs-length-comparison-NO-IO-8b}, \ref{fig:cost-vs-length-comparison-NO-IO-8c}, and \ref{fig:cost-vs-length-comparison-NO-IO-8d}, respectively.  The NSI parameters, the fundamental vacuum parameters, and the fixed neutrino energies for different experiments in the NO and IO scenarios, used in our analysis are taken from Tables \ref{tab:1}, \ref{tab:2}, \ref{tab:4}, and \ref{tab:5}, respectively.}
    \label{fig:cost-function-versus-length-NO-I0-comparison-8}
\end{figure}
\begin{figure}[H]
    \centering
    \begin{subfigure}[b]{0.34\textwidth}
        \centering
    \includegraphics[width=\textwidth]{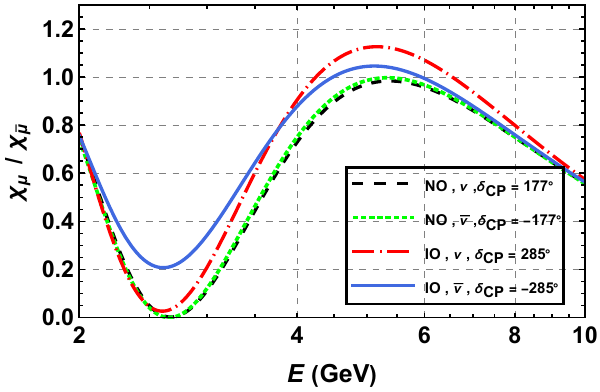}
        \caption{P2O}
        \label{fig:cost-vs-energy-NO-I0-Comparison-neutrino-antineutrino-9a}
    \end{subfigure}
    \hfill
    \begin{subfigure}[b]{0.34\textwidth}
        \centering
        \includegraphics[width=\textwidth]{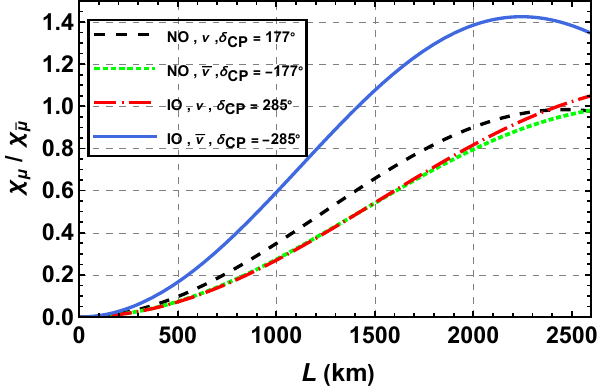}
        \caption{P2O}
        \label{fig:cost-vs-length-NO-I0-Comparison-neutrino-antineutrino-9b}
    \end{subfigure}
    \caption{We depict the cost function $\chi_\mu$ for the initial muon neutrino flavor state $\ket{\nu_\mu}$ (with positive matter potential $+V_{\rm CC}$ and positive $CP$-violation phase $+\delta_{\text{CP}}$) and $\chi_{\overline{\mu}}$ for the muon antineutrino flavor state $\ket{\overline{\nu}_\mu}$ (with negative matter potential $-V_{\rm CC}$ and negative $CP$-violation phase $-\delta_{\text{CP}}$) as functions of energy $E\,(\text{GeV})$ and Length $L\,(\text{km})$ in Figs.\,\ref{fig:cost-vs-energy-NO-I0-Comparison-neutrino-antineutrino-9a} and \ref{fig:cost-vs-length-NO-I0-Comparison-neutrino-antineutrino-9b}, respectively, in presence of NSI parameters, incorporating the fundamental vacuum parameters and a constant matter potential. The comparison is shown between NO and IO scenarios, along with their respective best-fit values of $\delta_{\text{CP}}$ and using the neutrino energy and baseline length of the P2O experiment. The NSI parameters, the fundamental vacuum parameters for the NO and IO scenarios, the baseline length and the fixed energy for the P2O neutrino experiment used in our analysis are taken from Tables\,\ref{tab:1}, \ref{tab:2}, \ref{tab:3}, and \ref{tab:6}, respectively.}
    \label{fig:cost-function-P2O-comp}
\end{figure}

Moreover, in Fig.\,\ref{fig:cost-function-versus-energy-NO-I0-comparison-7}, we examine the behavior of the cost function $\chi_\mu$ as a function of energy $E\,(\text{GeV})$ for the initial muon flavor neutrino state $\ket{\nu_\mu}$, comparing the NO and IO scenarios in the presence of NSI parameters together with the contributions from the fundamental vacuum parameters and a constant matter potential, along with the corresponding best fit values of $\delta_{\text{CP}}$. The comparisons are made by fixing the baseline lengths to those of the T2K, NOvA, DUNE, and P2O experiments, as shown in Figs.\,\ref{fig:cost-vs-energy-comparison-NO-IO-7a}, \ref{fig:cost-vs-energy-comparison-NO-IO-7b}, \ref{fig:cost-vs-energy-comparison-NO-IO-7c}, and \ref{fig:cost-vs-energy-comparison-NO-IO-7d}, respectively. We find that across all experiments, in the presence of real NSI parameters, the cost function satisfies $\chi_\mu = 1$ at $\delta_{\text{CP}} = 177^\circ$ (blue solid line) under the NO scenario. In contrast, under the IO scenario, $\chi_\mu > 1$ is consistently observed at $\delta_{\text{CP}} = 285^\circ$ (red dotted line). This indicates that, in the presence of NSI, the cost function is generally higher in the IO scenario at the best-fit value of $\delta_{\text{CP}}$ across all the considered experiments.

In Fig.\,\ref{fig:cost-function-versus-length-NO-I0-comparison-8}, we present a comparison of the cost function $\chi_\mu$ as a function of propagation length $L\,(\text{km})$ between the NO and IO scenarios for the initial muon neutrino flavor state $\ket{\nu_\mu}$. This analysis includes NSI parameters, fundamental vacuum parameters, a constant matter potential, and the corresponding best-fit values of $\delta_{\text{CP}}$. This comparison is made using the fixed neutrino energies listed in Tables\,\ref{tab:4} and \ref{tab:5} for the NSI cases under both NO and IO scenarios, corresponding to the T2K, NOvA, DUNE, and P2O experiments. As shown in Figs.\,\ref{fig:cost-vs-length-comparison-NO-IO-8a}, \ref{fig:cost-vs-length-comparison-NO-IO-8b}, \ref{fig:cost-vs-length-comparison-NO-IO-8c}, and \ref{fig:cost-vs-length-comparison-NO-IO-8d}, our finding further reinforces the general trend that the cost function is consistently larger in the IO scenario (red dotted line) compared to the NO scenario (blue solid line) in the presence of NSI, across all the experiments considered.

Additionally, we examine the cost function $\chi_{\overline{\mu}}$ for the initial muon antineutrino flavor state $\ket{\overline{\nu}_\mu}$. Fig.\,\ref{fig:cost-function-P2O-comp} presents both $\chi_\mu$ and $\chi_{\overline{\mu}}$, corresponding to the initial muon neutrino flavor state  $\ket{\nu_\mu}$ and the muon antineutrino flavor state $\ket{\overline{\nu}_\mu}$, respectively. These cost functions are shown  as functions of energy $E\,(\text{GeV})$ and baseline length $L\,(\text{km})$ in Figs.\,\ref{fig:cost-vs-energy-NO-I0-Comparison-neutrino-antineutrino-9a} and \ref{fig:cost-vs-length-NO-I0-Comparison-neutrino-antineutrino-9b}, respectively. The analysis includes the effects of real NSI parameters, fundamental vacuum parameters, and a constant matter potential under both the NO and IO scenarios using their respective best-fit $\delta_{\text{CP}}$ phase values, and employs the fixed baseline length and neutrino energy specific to the P2O experiment. In  Fig.\,\ref{fig:cost-vs-energy-NO-I0-Comparison-neutrino-antineutrino-9a}, we observe $\chi_\mu>1$ for the initial state $\ket{\nu_\mu}$ in the IO scenario with $\delta_{\text{CP}}=285^\circ$ (red dot-dashed line), and  $\chi_{\overline{\mu}}>1$ for the initial state $\ket{\overline{\nu}_\mu}$ in the IO scenario with $\delta_{\text{CP}}=-285^\circ$ (blue solid line). Likewise, Fig.\,\ref{fig:cost-vs-length-NO-I0-Comparison-neutrino-antineutrino-9b}, shows that at the end of the baseline length of P2O experiment, $\chi_\mu>1$ and $\chi_{\overline{\mu}}>1$ for both initial states $\ket{\nu_\mu}$ (red dot-dashed line) and $\ket{\overline{\nu}_\mu}$ (blue solid line), respectively, in the IO scenario with their respective best-fit $\delta_{\text{CP}}$ phase values. However, in the IO scenario, the cost function is higher for the initial muon antineutrino flavor state $\ket{\overline{\nu}_\mu}$ (blue solid line) compared to the initial muon neutrino flavor state $\ket{\nu_\mu}$ (red dot-dashed line). These findings further confirm that the cost function increases in the IO scenario when real NSI parameters and best-fit $\delta_{\text{CP}}$ values are considered in the calculation.

\section{Discussion and conclusion}
\label{Sec5}
We have employed the quantum spread complexity as a primary analytical tool to investigate three-flavor neutrino oscillations in vacuum and matter, incorporating the best-fit $CP$-violation phase ($\delta_{\text{CP}}$) and Non-Standard Interaction (NSI) effects. Our analysis considered two neutrino mass ordering scenarios: normal ordering (NO) and inverted ordering (IO). A comprehensive set of cases has been examined for each mass ordering, including vacuum oscillations, oscillations in a constant matter potential, and those involving NSI real parameters, using both $\delta_{\text{CP}}=0$ and the best-fit values ($\delta_{\text{CP}}=177^\circ$ for NO and $\delta_{\text{CP}}=285^\circ$ for IO). 

At first, the transition probabilities have been computed for both the initial muon neutrino and muon antineutrino flavor states. These transition probabilities, specifically $P_{\mu\rightarrow e}$, $P_{\mu\rightarrow \mu}$, and $P_{\mu\rightarrow \tau}$, along with their CP-conjugate counterparts, have been evaluated separately as functions of energy and length. Our analysis used baseline lengths and energy ranges corresponding to the currently running long-baseline accelerator experiments such as T2K and NOvA, and the upcoming experiments DUNE and P2O. For each set of cases in the NO and IO scenarios, we have identified the energy at which the first oscillation $P_{\mu\rightarrow e}$ maximum occurred and then fixed this energy to further analyze the transition probabilities as a function of the propagation length. Discrepancies in transition probabilities were observed across different oscillation scenarios for both the initial muon neutrino and muon antineutrino flavor states, particularly at the end of the baseline for each experiment.

To probe the complexity of the underlying neutrino quantum evolution, we have applied the concept of quantum spread complexity in neutrino oscillations, which can be quantified by the cost function. We have derived this cost function from the constructed Krylov basis of the initial muon neutrino flavor states, which could be expressed in terms of observable flavor transition probabilities. This has served as a diagnostic tool for the spread of the quantum state in flavor space. Using the fixed baseline lengths of the T2K, NOvA, DUNE, and P2O experiments and analyzing the cost function as a function of energy, we have observed that in the NO scenario, discrepancies between different oscillation cases were minimal for the T2K, NOvA, and DUNE experiments but became more substantial in the P2O configuration. In contrast, the IO scenario exhibited persistent discrepancies in the cost function across all experiments. Particularly striking was the observation that the cost function exceeded unity for the initial muon neutrino flavor state in the IO scenario at the best-fit $\delta_{\text{CP}}=285^\circ$, when the NSI real parameters with fundamental vacuum parameters and a constant matter potential contributions were included. This behavior indicates a greater spread of the quantum state over the flavor basis, which implies a higher complexity of the neutrino system evolution in IO compared to NO. We further examined the cost function as a function of propagation length, using the energy corresponding to the first oscillation peak for each case under both NO and IO scenarios. The IO scenario consistently yielded higher cost function values at the end of the baseline length of all experiments. These results can be primarily ascribed to the transition probabilities $P_{\mu\rightarrow e}$, $P_{\mu \rightarrow \tau}$, and the interference term appeared in the cost function expression.

Moreover, we have extended our analysis to the initial muon antineutrino flavor state, using the baseline length and energy range of the P2O experiment. Under the same physical conditions, presence of real NSI parameters, a constant matter potential, fundamental vacuum parameters, and the best-fit $CP$-violation phase in IO, the cost function again exceeded unity, reaching values as high as approximately 1.4 at the end of the baseline of the P2O experiment. This result underscores an even greater spread in the antineutrino sector and highlights the NSI effects, $CP$ violation, and mass ordering-dependent nature of the cost function. From a physical perspective, quantum spread complexity characterizes how an initial state spreads in the Hilbert space under unitary evolution. A cost function value of 1 corresponds to a quantum state that spreads exactly over the three-dimensional Hilbert space of the standard neutrino system. Values exceeding 1, seen in the IO scenario, indicate the spread extends beyond this space, potentially hinting at the involvement of additional degrees of freedom. Although sterile neutrinos have not yet been detected in neutrino experiments, our results raise the intriguing possibility that additional basis states, such as those associated with sterile neutrinos, may be required to fully describe the dynamics in scenarios that exhibit a large cost function. Alternatively, from the complexity studies made in this work, it could be suggested that NO is preferred by nature over IO in the context of neutrino oscillations.

Our findings have implications for neutrino oscillation experiments. The sensitivity of the cost function to different experiments such as T2K, NOvA, DUNE, and P2O, during neutrino propagation in matter, in the presence of NSI effects, $CP$ violation, and mass ordering scenarios, suggests that these tools can effectively complement traditional analyses based solely on transition probabilities. In particular, the enhanced cost function observed in the IO scenario at the best-fit $\delta_{\text{CP}}$ values and in the presence of NSI provides a potential signature for distinguishing between mass ordering scenarios and probing new physics beyond the Standard Model.

In conclusion, we have demonstrated that quantum spread complexity offers a novel and informative perspective on neutrino oscillation phenomena. It could capture not only the probabilistic features of flavor transition but also the underlying structural evolution of the quantum state. Our results indicate that this approach could aid in unraveling the nature of NSI effects, the $CP$-violation phase, and mass ordering scenarios in neutrino oscillations in matter, particularly in high-precision long-baseline accelerator neutrino experiments such as T2K, NOvA, DUNE, and P2O.

\section*{Acknowledgments}
A.K.J. and S.B. would like to acknowledge the project funded by SERB, India, with Ref. No. CRG/2022/003460, for supporting this research. We would like to dedicate this work to the memory of Ashutosh K. Alok.

\appendix
\section{Appendix}
\label{Append}
\noindent
\resizebox{\textwidth}{!}{%
\begin{minipage}{\textwidth}
\begin{equation*}
a_1 =  V_{CC} \varepsilon_{e\mu} + \frac{\Delta m_{21}^2 \, U_{e2} \, U_{\mu 2}^* + \Delta m_{31}^2 \, U_{e3} \, U_{\mu 3}^*}{2E}
\end{equation*}
\end{minipage}
}\\

\noindent
\resizebox{\textwidth}{!}{%
\begin{minipage}{\textwidth}
\begin{equation*}
a_2 =  V_{CC} \varepsilon_{\tau \mu} + \frac{\Delta m_{21}^2 \, U_{\tau 2} \, U_{\mu 2}^* + \Delta m_{31}^2 \, U_{\tau 3} \, U_{\mu 3}^*}{2E}
\end{equation*}
\end{minipage}
}\\

\noindent
\resizebox{\textwidth}{!}{%
\begin{minipage}{\textwidth}
\begin{align*}
&b_1 = \left( V_{CC} \varepsilon_{ee} + V_{CC} + \frac{\Delta m_{21}^2 U_{e2} U_{e2}^* + \Delta m_{31}^2 U_{e3} U_{e3}^*}{2E} \right) 
\left( V_{CC} \varepsilon_{e\mu} + \frac{\Delta m_{21}^2 U_{e2} U_{\mu 2}^* + \Delta m_{31}^2 U_{e3} U_{\mu 3}^*}{2E} \right) \\
&+ \left( V_{CC} \varepsilon_{\mu \mu} + \frac{\Delta m_{21}^2 U_{\mu 2} U_{\mu 2}^* + \Delta m_{31}^2 U_{\mu 3} U_{\mu 3}^*}{2E} \right)
\left( V_{CC} \varepsilon_{e\mu} + \frac{\Delta m_{21}^2 U_{e2} U_{\mu 2}^* + \Delta m_{31}^2 U_{e3} U_{\mu 3}^*}{2E} \right) \\
&+ \left(
  V_{CC}\,\varepsilon_{\tau\mu} + 
  \frac{
    \Delta m_{21}^2U_{t2}U^\star_{m2} + 
    \Delta m_{31}^2U_{t3}U^\star_{m3}
  }{2E}
\right)
\left(
  V_{CC}\,\varepsilon_{e\tau} + 
  \frac{
    \Delta m_{21}^2U_{e2}U^\star_{t2} + 
    \Delta m_{31}^2U_{e3}U^\star_{t3}
  }{2E}
\right) \\
&- \frac{1}{
   \abs{ \left( V_{CC} \varepsilon_{e\mu} + \frac{\Delta m_{21}^2 U_{e2} U_{\mu 2}^* + \Delta m_{31}^2 U_{e3} U_{\mu 3}^*}{2E} \right)}^2
   + \abs{\left( V_{CC} \varepsilon_{\tau \mu} + \frac{\Delta m_{21}^2 U_{\tau 2} U_{\mu 2}^* + \Delta m_{31}^2 U_{\tau 3} U_{\mu 3}^*}{2E} \right)}^2
} \\
&\times \left\{
\left( V_{CC} \varepsilon_{e\mu} + \frac{\Delta m_{21}^2 U_{e2} U_{\mu 2}^* + \Delta m_{31}^2 U_{e3} U_{\mu 3}^*}{2E} \right)^* \right. \\
&\quad \times \left[
\left( V_{CC} \varepsilon_{ee} + V_{CC} + \frac{\Delta m_{21}^2 U_{e2} U_{e2}^* + \Delta m_{31}^2 U_{e3} U_{e3}^*}{2E} \right)
\left( V_{CC} \varepsilon_{e\mu} + \frac{\Delta m_{21}^2 U_{e2} U_{\mu 2}^* + \Delta m_{31}^2 U_{e3} U_{\mu 3}^*}{2E} \right) \right. \\
&\quad + \left( V_{CC} \varepsilon_{\mu \mu} + \frac{\Delta m_{21}^2 U_{\mu 2} U_{\mu 2}^* + \Delta m_{31}^2 U_{\mu 3} U_{\mu 3}^*}{2E} \right)
\left( V_{CC} \varepsilon_{e\mu} + \frac{\Delta m_{21}^2 U_{e2} U_{\mu 2}^* + \Delta m_{31}^2 U_{e3} U_{\mu 3}^*}{2E} \right) \\
&\quad + \left( V_{CC} \varepsilon_{e\tau} + \frac{\Delta m_{21}^2 U_{e2} U_{\tau 2}^* + \Delta m_{31}^2 U_{e3} U_{\tau 3}^*}{2E} \right)
\left( V_{CC} \varepsilon_{\tau \mu} + \frac{\Delta m_{21}^2 U_{\tau 2} U_{\mu 2}^* + \Delta m_{31}^2 U_{\tau 3} U_{\mu 3}^*}{2E} \right)
\Bigg] \\
&\quad + \left( V_{CC} \varepsilon_{\tau \mu} + \frac{\Delta m_{21}^2 U_{\tau 2} U_{\mu 2}^* + \Delta m_{31}^2 U_{\tau 3} U_{\mu 3}^*}{2E} \right)^* \\
&\quad \times \left[
\left( V_{CC} \varepsilon_{e\mu} + \frac{\Delta m_{21}^2 U_{e2} U_{\mu 2}^* + \Delta m_{31}^2 U_{e3} U_{\mu 3}^*}{2E} \right)
\left( V_{CC} \varepsilon_{\tau e} + \frac{\Delta m_{21}^2 U_{\tau 2} U_{e2}^* + \Delta m_{31}^2 U_{\tau 3} U_{e3}^*}{2E} \right) \right. \\
&\quad + \left( V_{CC} \varepsilon_{\mu \mu} + \frac{\Delta m_{21}^2 U_{\mu 2} U_{\mu 2}^* + \Delta m_{31}^2 U_{\mu 3} U_{\mu 3}^*}{2E} \right)
\left( V_{CC} \varepsilon_{\tau \mu} + \frac{\Delta m_{21}^2 U_{\tau 2} U_{\mu 2}^* + \Delta m_{31}^2 U_{\tau 3} U_{\mu 3}^*}{2E} \right) \\
&\quad + \left. \left( V_{CC} \varepsilon_{\tau \tau} + \frac{\Delta m_{21}^2 U_{\tau 2} U_{\tau 2}^* + \Delta m_{31}^2 U_{\tau 3} U_{\tau 3}^*}{2E} \right)
\left( V_{CC} \varepsilon_{\tau \mu} + \frac{\Delta m_{21}^2 U_{\tau 2} U_{\mu 2}^* + \Delta m_{31}^2 U_{\tau 3} U_{\mu 3}^*}{2E} \right)
\right]
\Bigg\}
\end{align*}
\end{minipage}
}\\

\begin{align*}
b_2 = &\left( V_{CC} \varepsilon_{\tau e} + \frac{ \Delta m_{21}^2 \, U_{\tau 2} \, U_{e2}^* + \Delta m_{31}^2 \, U_{\tau 3} \, U_{e3}^* }{2E} \right)
\left( V_{CC} \varepsilon_{e\mu} + \frac{ \Delta m_{21}^2 \, U_{e2} \, U_{\mu 2}^* + \Delta m_{31}^2 \, U_{e3} \, U_{\mu 3}^* }{2E} \right) \\
&+ \left( V_{CC} \varepsilon_{\mu \mu} + \frac{ \Delta m_{21}^2 \, U_{\mu 2} \, U_{\mu 2}^* + \Delta m_{31}^2 \, U_{\mu 3} \, U_{\mu 3}^* }{2E} \right)
\left( V_{CC} \varepsilon_{\tau \mu} + \frac{ \Delta m_{21}^2 \, U_{\tau 2} \, U_{\mu 2}^* + \Delta m_{31}^2 \, U_{\tau 3} \, U_{\mu 3}^* }{2E} \right) \\
&+ \left( V_{CC} \varepsilon_{\tau \mu} + \frac{ \Delta m_{21}^2 \, U_{\tau 2} \, U_{\mu 2}^* + \Delta m_{31}^2 \, U_{\tau 3} \, U_{\mu 3}^* }{2E} \right)
\left( V_{CC} \varepsilon_{\tau \tau} + \frac{ \Delta m_{21}^2 \, U_{\tau 2} \, U_{\tau 2}^* + \Delta m_{31}^2 \, U_{\tau 3} \, U_{\tau 3}^* }{2E} \right) \\
&- \frac{1}{
    \left| V_{CC} \varepsilon_{e\mu} + \frac{\Delta m_{21}^2 U_{e2} U_{\mu2}^* + \Delta m_{31}^2 U_{e3} U_{\mu3}^*}{2E} \right|^2 +
    \left| V_{CC} \varepsilon_{\tau\mu} + \frac{\Delta m_{21}^2 U_{\tau2} U_{\mu2}^* + \Delta m_{31}^2 U_{\tau3} U_{\mu3}^*}{2E} \right|^2
}\\
& \times \Bigg \{ \left( V_{CC} \varepsilon_{\tau\mu} + \frac{\Delta m_{21}^2 U_{\tau2} U_{\mu2}^* + \Delta m_{31}^2 U_{\tau3} U_{\mu3}^*}{2E} \right)
\cdot  \Bigg(\left( V_{CC} \varepsilon_{e\mu} + \frac{\Delta m_{21}^2 U_{e2} U_{\mu2}^* + \Delta m_{31}^2 U_{e3} U_{\mu3}^*}{2E} \right)^* \\
&\quad \times \Bigg[
\left( V_{CC} \varepsilon_{ee} + V_{CC} + \frac{\Delta m_{21}^2 |U_{e2}|^2 + \Delta m_{31}^2 |U_{e3}|^2}{2E} \right)
\left( V_{CC} \varepsilon_{e\mu} + \frac{\Delta m_{21}^2 U_{e2} U_{\mu2}^* + \Delta m_{31}^2 U_{e3} U_{\mu3}^*}{2E} \right) \\
&\quad + \left( V_{CC} \varepsilon_{\mu\mu} + \frac{\Delta m_{21}^2 |U_{\mu2}|^2 + \Delta m_{31}^2 |U_{\mu3}|^2}{2E} \right)
\left( V_{CC} \varepsilon_{e\mu} + \frac{\Delta m_{21}^2 U_{e2} U_{\mu2}^* + \Delta m_{31}^2 U_{e3} U_{\mu3}^*}{2E} \right) \\
&\quad + \left( V_{CC} \varepsilon_{e\tau} + \frac{\Delta m_{21}^2 U_{e2} U_{\tau2}^* + \Delta m_{31}^2 U_{e3} U_{\tau3}^*}{2E} \right)
\left( V_{CC} \varepsilon_{\tau\mu} + \frac{\Delta m_{21}^2 U_{\tau2} U_{\mu2}^* + \Delta m_{31}^2 U_{\tau3} U_{\mu3}^*}{2E} \right)
\Bigg] \\
&\quad + \left( V_{CC} \varepsilon_{\tau\mu} + \frac{\Delta m_{21}^2 U_{\tau2} U_{\mu2}^* + \Delta m_{31}^2 U_{\tau3} U_{\mu3}^*}{2E} \right)^* \\
&\quad \times \Bigg[
\left( V_{CC} \varepsilon_{e\mu} + \frac{\Delta m_{21}^2 U_{e2} U_{\mu2}^* + \Delta m_{31}^2 U_{e3} U_{\mu3}^*}{2E} \right)
\left( V_{CC} \varepsilon_{\tau e} + \frac{\Delta m_{21}^2 U_{\tau2} U_{e2}^* + \Delta m_{31}^2 U_{\tau3} U_{e3}^*}{2E} \right) \\
&\quad + \left( V_{CC} \varepsilon_{\mu\mu} + \frac{\Delta m_{21}^2 |U_{\mu2}|^2 + \Delta m_{31}^2 |U_{\mu3}|^2}{2E} \right)
\left( V_{CC} \varepsilon_{\tau\mu} + \frac{\Delta m_{21}^2 U_{\tau2} U_{\mu2}^* + \Delta m_{31}^2 U_{\tau3} U_{\mu3}^*}{2E} \right) \\
&\quad + \left( V_{CC} \varepsilon_{\tau\tau} + \frac{\Delta m_{21}^2 |U_{\tau2}|^2 + \Delta m_{31}^2 |U_{\tau3}|^2}{2E} \right)
\left( V_{CC} \varepsilon_{\tau\mu} + \frac{\Delta m_{21}^2 U_{\tau2} U_{\mu2}^* + \Delta m_{31}^2 U_{\tau3} U_{\mu3}^*}{2E} \right)
\Bigg]
\Bigg)
\Bigg\}
\end{align*}


\bibliography{references}
\end{document}